\documentclass[11pt]{article}
\pdfoutput=1
\usepackage{amsmath}
\usepackage{amssymb}
\usepackage{amsfonts}
\usepackage{mathrsfs}
\usepackage{mathrsfs}
\usepackage{fullpage}
\usepackage{setspace}
\usepackage{bm}
\usepackage{bbm}
\usepackage{relsize}
\usepackage{adjustbox}
\usepackage{multirow}
\usepackage{cite}
\usepackage{filecontents}
\usepackage{scalerel}
\usepackage[normalem]{ulem}
\usepackage{enumerate}
\usepackage{yfonts}
\usepackage{psfrag}
\usepackage{array}
\usepackage{booktabs}
\newcolumntype{N}{>{\centering\arraybackslash}m{.5in}}
\newcolumntype{G}{>{\centering\arraybackslash}m{2in}}
\usepackage[symbol]{footmisc}

\usepackage[latin1]{inputenc}
\usepackage{graphicx}
\usepackage{cancel}
\usepackage{slashed}
\usepackage{mathtools}

\usepackage[font=small,labelfont=bf]{caption}
\usepackage{subcaption}

\usepackage[colorlinks=true]{hyperref}
\hypersetup{
    unicode=false,          
    pdftoolbar=true,        
    pdfmenubar=true,        
    pdffitwindow=false,     
    pdfstartview={FitH},    
    pdftitle={Gravitational waves from regular black holes in extreme mass-ratio inspirals},    
    pdfauthor={Shailesh Kumar, Tieguang Zi},     
    pdfnewwindow=true,      
    colorlinks=true,       
    linkcolor=blue,          
    citecolor=red,        
    filecolor=magenta,      
    urlcolor=blue           
    }

\usepackage{hyperref}
\hypersetup{colorlinks=true}

\def\equationautorefname~#1\null{%
	Eq.~(#1)\null
}
\def\figureautorefname~#1\null{%
	Fig.~#1\null
}
\def\tableautorefname~#1\null{%
	Table.~#1\null
}
\def\sectionautorefname~#1\null{%
	Section #1\null
}
\def\appendixautorefname~#1\null{%
	Appendix #1\null
}

\begin{document}

\numberwithin{equation}{section}
{
\begin{titlepage}
\begin{center}

\hfill \\
\hfill \\
\vskip 0.75in

{\Large {\bf Gravitational waves from regular black holes in \\ extreme mass-ratio inspirals
}}

\vskip 0.24in


{\large Shailesh Kumar${}$$^{a}$, Tieguang Zi${}$\footnote[2]{\href{mailto: zitieguang@ncu.edu.cn}{zitieguang@ncu.edu.cn}}}$^{b,c,d}$\vspace{0.25in}

{\it ${}$$^a$Indian Institute of Technology, Gandhinagar, Gujarat-382355, India \vspace{0.12cm}\\}

{\it ${}$$^b$Department of physics, Nanchang University, Nanchang, 330031, China}

{\it ${}$$^c$Center for Relativistic Astrophysics and High Energy Physics, Nanchang University, Nanchang, 330031, China}

{\it ${}$$^d$School of Physics and Optoelectronics, South China University of Technology, Guangzhou 510641, People's Republic of China}


\vskip.5mm

\end{center}

\vskip 0.35in

\begin{center}
{\bf ABSTRACT }
\end{center}
We analyze a rotating regular black hole spacetime with an asymptotically Minkowski core, focusing on extreme mass-ratio inspiral (EMRIs) where a stellar-mass object inspirals a supermassive black hole under consideration. Such spacetimes are also called Kerr-like spacetimes, which motivate the investigation of black holes beyond general relativity and the test of the no-hair theorem. In the present article, we consider the eccentric equatorial motion of an inspiralling object in the background of a rotating regular black hole. The dynamics generate gravitational waves (GWs) that imply a loss in energy and angular momentum of the orbiting body. In this scenario, with a slow-rotation approximation and the radiation reaction, we analytically compute the orbital evolution of the moving object. Further, we generate the gravitational waveforms and constrain the non-Kerr parameter through dephasing and mismatch computations using Laser Interferometer Space Antenna (LISA) observations. Our result indicates that LISA can distinguish the effect of the additional non-Kerr/deviation parameter with the parameter as small as $\sim10^{-6}$. The constraint on the parameter in the regular black hole using the Fisher information matrix (FIM) can be obtained within a fraction error of $10^{-5}$. The estimates of our analysis with EMRIs present the possible detectability of Kerr-like geometries with future space-based detectors and further open up ways to put a stringent constraint on non-Kerr parameters with more advanced frameworks.
\end{titlepage}
}

\newpage
\tableofcontents

\section{Introduction}

Black holes, a mysterious entity that has fascinated scientists to present major theoretical, observational, and mathematical hurdles, are regions of spacetime whose solution first appeared under the Einstein's theory of general relativity (GR). After almost a century, as researchers continue to investigate the underlying nature of such objects and their place in the cosmos, these bodies have led a great deal of study that has produced advances in understanding the fundamental aspects of gravitational physics from theoretical and observational perspectives. As per the current status, GR has been widely tested and proven successful theory across different scales \cite{Will:2014kxa}, providing new avenues for analysing distinct characteristics of strong and weak regimes of gravity. In GR, spacetime singularities are inescapable and arise in most physically meaningful solutions to Einstein field equations, particularly exact solutions such as Schwarzschild and Kerr black holes \cite{Hawking:1973uf}. Also, astrophysical black holes can be accurately modelled using the Kerr metric and are believed to reside at the centre of galaxies. This notion, termed the Kerr hypothesis, relies on the uniqueness theorem, which supports the no-hair theorem in GR, asserting that the Kerr and Kerr-Newman solutions are the sole stationary, axisymmetric, and asymptotically flat solutions to Einstein's field equations in vacuum and electro-vacuum \cite{PhysRev.164.1776, PhysRev.174.1559, Hawking1972, PhysRevLett.34.905}. As such solutions do not have clear and direct observational evidence, it is difficult to rule out the existence of non-Kerr black holes \cite{Will2006, Bambi_2011}. The GW community is actively working to provide effective tests to identify possible deviations from GR predictions in GW data. Despite the absence of deviations from GR in current GW data \cite{LIGOScientific:2019fpa, LIGOScientific:2020tif, LIGOScientific:2021sio}, ongoing advancements in observational astrophysics present an opportunity to evaluate the validity of the no-hair theorem and conduct a more comprehensive investigation of strong and weak gravity regimes of black holes.

An important milestone in developing the early stages of mathematical relativity was the Penrose-Hawking singularity theorems that remain essential to the framework of the current gravitational research \cite{PhysRevLett.14.57, Hawking:1973uf, Bueno:2024zsx}. These theorems meticulously show that singularities are a necessary consequence of gravitational collapse. It also indicated that singularities are places where our existing physical theories fall short, suggesting a fundamental failure in our comprehension of the natural laws. In the event that singularities exist, our capacity to model reality would be severely challenged since they would add an unsolvable indeterminacy into our description of reality of the cosmos. In view of this, it is generally expected that quantum gravitational effects will solve singularities by examining underlying phenomena at small scales and providing a more thorough explanation of gravity under extreme circumstances. The behaviour of matter is affected by quantum phenomena at very high densities. As matter approaches the Planck density, these quantum gravitational forces may provide enough counteracting pressure to avoid the formation of a singularity \cite{Lopez:2018aec, Jafarzade:2021umv}. The study of non-singular black holes has been driven by this notion, which has inspired research into black holes that avoid singularities.

Bardeen was one of the pioneers in developing a non-singular black hole model, replacing the singularity with a de Sitter core \cite{1968qtr..conf...87B}, that involved coupling GR with an electromagnetic field. Since then, numerous black hole models with regular cores have been proposed \cite{Bronnikov:2000vy, Dymnikova:2004zc, Balart:2014jia, Neves:2014aba, Balart:2014cga, Lobo:2020ffi} with various viewpoints of analysis \cite{Myung:2007xd, Carballo-Rubio:2021bpr}, including rotating regular black holes and their properties \cite{Ghosh:2014hea, Eichhorn:2021etc}. A regular black hole contains a horizon, finite curvature tensors and their invariance. It is imperative from a strategic perspective to apply physical constraints that are directly related to the goals and objectives of the observational community. If we concentrate on the most effective or ideal spacetime candidate, it should meet specific requirements. For example, it should be axisymmetric, asymptotically flat, separability of the Hamilton-Jacobi equations and constraints on classical energy conditions, etc. One can further impose other constraints, such as the separability of the Dirac equation and closed timelike curves, etc \cite{Simpson:2021zfl}. Such a list of physically motivated individuals should be refreshed to include the most current or forthcoming information. It is intricate to construct such geometries which fulfill these physically motivated constraints and further motivate researchers to focus in this direction. Recently, there has been much interest in a novel and intriguing model of a regular black hole provided by Simpson-Visser where the spacetime geometry has an asymptotically Minkowski core \cite{Simpson:2019mud, Simpson:2021zfl}, which remains an interest of the present article. This regular geometry was independently developed by Ghosh \cite{Ghosh:2014pba}, and further more comprehensively by Simpson and Visser \cite{Simpson:2021zfl}. Since spin plays a key role in determining how black holes behave in astrophysical settings, non-rotating black holes might not be evaluated using astrophysical observations. Given this, we consider the stationary, axisymmetric and non-singular/regular rotating black hole which is described by three parameters: mass ($M$), spin ($a$) and a free parameter ($\ell$) that implies the potential deviation from the Kerr geometry \cite{Simpson:2021zfl}, also termed as non-Kerr/deviation parameter. It also holds the Killing tower of a Killing tensor \cite{Frolov:2017kze} that indicates the existence of a Carter constant \cite{PhysRev.174.1559}, further implying the separability of the Hamilton-Jacobi equation. We often term this black hole geometry as Kerr-like spacetime.

The present article examines the observational consequences of a rotating regular black hole in the context of extreme mass-ratio inspirals (EMRIs), where a stellar-mass object (called secondary) inspirals a supermassive black hole (called primary), maintaining a mass-ratio $(q)$ in the range of ($10^{-4} - 10^{-7}$). In particular, we consider a point particle object hovering in the vicinity of a rotating regular black hole spacetime. The inspiral motion of the secondary in the background of the primary generates GWs. Such an event induces a backreaction on the object, and as a result, it truncates its trajectory near the last stable orbits (LSO). This inspiral process is crucial for probing various characteristics of weak and strong regimes of gravity through distinct methodologies such as black hole perturbation techniques and post-Newtonian (PN) frameworks \cite{Pound:2021qin, Blanchet:2013haa}. In this respect, EMRIs will be useful in unfolding the fundamental facets of gravity theory based on future observations with space-based detectors like Laser Interferometer Space Antenna (LISA) and TianQin \cite{amaroseoane2017laserinterferometerspaceantenna, TianQin:2015yph, TianQin:2020hid, Cardenas-Avendano:2024mqp, PhysRevD.95.103012}. It can quantify the parameters of black holes and investigate foundational aspects of gravity with unprecedented accuracy \cite{Barack:2006pq, Babak:2017tow, Fan:2020zhy, Zi:2021pdp}. There is plenty of literature available in this line of interest to test GR and beyond as well as to probe distinct features of strong and weak gravity regimes of black holes \cite{Kocsis:2011dr, Barausse:2014tra, Cardoso:2022whc, Figueiredo:2023gas, Rahman:2023sof, Rahman:2022fay, Zhang:2024ugv, Duque:2024mfw, PhysRevLett.129.241103, PhysRevD.105.L061501, PhysRevLett.126.141102, Destounis:2022obl, Kumar:2024utz, AbhishekChowdhuri:2023gvu, Datta:2024vll, Zi:2023pvl, Zi:2024jla, Kumar:2024dql, Fu:2024cfk, Zhang:2024csc, Loutrel:2024qxp, Zhang:2024ogc, Zi:2024mbd, Zi:2024dpi, Zi:2023qfk, Zi:2023omh, Zi:2022hcc}. We implement particularly leading-order PN \cite{Moore:2016qxz, Blanchet:2013haa, Ryan:1995xi, PhysRevD.50.3816} analysis to examine the impact of the non-Kerr/deviation parameter in the GW waveforms for the rotating regular black holes in EMRIs, along with the constraints on the deviation to the Kerr geometry with the LISA mission.

Let us take a look at how we organize the draft. We start with a brief description of the rotating regular black hole geometry and the equatorial orbital motion of the inspiralling object in Section (\ref{BHspacetime}). In Section (\ref{radiation:reaction}), we provide basic ingredients for computing analytical results and further obtain the average rate of change of orbital energy and angular momentum under slow-rotation approximation, including the eccentric orbital dynamics of the inspiralling object. We next, in Sections (\ref{dphasewavefomr}) and (\ref{detect}), estimate dephasing, generate waveforms, mismatch, and finally use the Fisher Information Matrix (FIM) to constrain the deviation parameter for its detectability from LISA observations along with the measurement error of the source parameters. Lastly, we conclude with a discussion in Section (\ref{discussion}), giving a summary of the obtained results and future prospects. We also add the other computational details in Appendices (\ref{apenteu1}), (\ref{fim}) and (\ref{appC}). \vspace{0.2cm}

\par
\textit{Notation and Convention: } We set the fundamental constants $G$ and $c$ to unity and adopt the positive sign convention $(-1,1,1,1)$. Roman letters are used to denote spatial indices, and Greek letters are used to represent four-dimensional indices.

\section{Regular black hole and orbital dynamics} \label{BHspacetime}
In this section, we briefly discuss the rotating regular black hole geometry, for which we are interested in exploring the observational signatures of a non-Kerr parameter in GWs. In order to construct such a spacetime, a strategic approach to imposing physics constraints in ways that directly resonate with observational efforts enables clearer connections between theory and measurable phenomena. For example, a theoretically desirable and preferred black hole spacetime should comply with certain requirements such as axisymmetry, asymptotic flatness and separability of Hamilton-Jacobi equations, etc \cite{Simpson:2021zfl}. In light of this, Simpson and  Visser (SV) recently constructed a spacetime that models a tractable rotating regular black hole with an asymptotically Minkowski core and asymptotically Kerr at large distances \cite{Simpson:2021zfl, Simpson:2019mud, Ghosh:2014pba}, given by
\begin{align}\label{metric}
ds^{2} = -\frac{\Delta}{\Sigma}(dt-a\sin^{2}\theta d\phi)^{2}+\frac{\Sigma}{\Delta}dr^{2}+\Sigma d\theta^{2}+\frac{\sin^{2}\theta}{\Sigma}[(r^{2}+a^{2})d\phi-a dt]^{2}
\end{align}
where, $\Sigma = r^{2}+a^{2}\cos^{2}\theta$ and $\Delta = r^{2}+a^{2}-2M r e^{-\ell/r}$ with $\ell>0$. $M$ is the mass of the supermassive black hole, and $a$ is the spin of the black hole. The parameter $\ell$ can be viewed as the deviation from the Kerr geometry. The metric (\ref{metric}) is also known as the \textit{eye of the storm spacetime} or \textit{Kerr-like spacetime}. It adheres to all standard energy conditions in GR across regions deemed theoretically and observationally valid while also ensuring that geodesics are integrable \cite{Simpson:2021zfl}. One can easily see that the asymptotic flatness is maintained when $r\longrightarrow \infty$, as well as $\ell\longrightarrow 0$ recovers the Kerr spacetime. Also, if one sets $a\longrightarrow 0$ and keeps $\ell$ non-vanishing, one can obtain its spherically symmetric version \cite{Simpson:2019mud}. Therefore, the spacetime under consideration has been developed based on thoughtfully selected criteria grounded in physical principles and observational motivations. It has been demonstrated that when applying a typical general relativity analysis to the resulting spacetime, the metric represents a regular black hole in the sense of Bardeen \cite{1968qtr..conf...87B}. Thus the spacetime is astrophysically relevant and can serve as a useful tool for observational and GW astronomy perspectives \cite{Simpson:2019mud, Ghosh:2014pba}.

\subsection{Orbital dynamics}
Our analysis examines the orbital dynamics of an inspiralling object that exhibits eccentric equatorial motion in the background of the central supermassive black hole defined by Eq. \eqref{metric}. As the described spacetime is stationary and possesses axisymmetry, it holds two constants of motion energy and angular momentum ($E, J_{z}$), along with another constant ($\mathcal{Q}$) known as the Carter constant, which will turn to zero since we focus on equatorial motion \cite{Glampedakis:2002cb, Glampedakis:2002ya, PhysRevD.61.084004}. It is to note that we perform the computations by making the parameters dimensionless \cite{Kumar:2024utz, AbhishekChowdhuri:2023gvu}; however, we avoid giving any special notions for writing convenience. One can always express the computed results in physical units wherever required. Let us now mention the 4-velocity of the inspiralling object.

\begin{equation}
\begin{aligned}\label{geo}
\mu\frac{dt}{d\tau} =& \frac{1}{\Delta\Sigma}\Big[\left(a^2+r^2\right) \left(a^2 E-a J_{z}+E r^2\right)+a \Delta \left(J_{z}-a E \sin ^2\theta \right) \Big] \\
\mu\frac{d\phi}{d\tau} =& \frac{1}{\Delta\Sigma}\Big[a \left(a^2 E-a J_{z}+E r^2-E \Delta\right)+J_{z} \csc ^2\theta  \Delta\Big] \\
\mu^{2}\Sigma^{2}\Big(\frac{dr}{d\tau}\Big)^{2} =& \Big(E(a^{2}+r^{2})-aJ_{z}\Big)^{2}-\Delta(\kappa+\mu^{2}r^{2}) \\
\mu^{2}\Sigma^{2}\Big(\frac{d\theta}{d\tau}\Big)^{2} =& (\kappa - \mu^{2}a^{2}\cos^{2}\theta)-\Big(aE\sin\theta-\frac{J_{z}}{\sin\theta}\Big)^{2},
\end{aligned}
\end{equation}
where $\mathcal{Q}\equiv\kappa-(J_{z}-aE)^{2}$ and $\mu$ is the mass of the inspiralling object. Since we are dealing with equatorial consideration, $\mathcal{Q}=0$, as a results, $\kappa = J_{z}^{2}-2aEJ_{z}+\mathcal{O}(a^{2})$. Other relevant expressions have been mentioned in Appendix (\ref{apenteu1}). Taking the radial part of Eq. (\ref{geo}), one can determine the region where inspiralling object truncates its trajectory. Before we obtain such a quantity, we know that eccentric orbits exhibit two turning points, \textit{periastron} ($r_{p}$) and \textit{apastron} ($r_{a}$), i.e, $r_{p,a} = p/(1\pm e)$. Using the fact that the radial velocity vanishes at the turning points as well as the bounded orbits are obtained in the region $r_{p}<r<r_{a}$, we get

\begin{align}
E =& \frac{1}{p\sqrt{2}}\sqrt{\frac{\left(e^2-1\right)^2 \ell^2-4 \left(e^2-1\right)^2 \ell-8 e^2 p+2 (p-2)^2 p}{p-3-e^2}} -\frac{a\left(e^2-1\right)^2}{{\sqrt{2} p^2 \left(p-3-e^2\right)}}\sqrt{\frac{\left(e^2+3\right) \ell^2-4 \ell p+2 p^2}{p-3-e^2}} \\
J_{z} =& \frac{1}{\sqrt{2}}\sqrt{-\frac{\left(e^2+3\right) \ell^2-4 \ell p+2 p^2}{p-3-e^2}}-\frac{a\left(e^2+3\right)}{\sqrt{2} p \left(p-3-e^2\right)}\sqrt{\frac{\left(e^2-1\right)^2 \ell^2-4 \left(e^2-1\right)^2 \ell-8 e^2 p+2 (p-2)^2 p}{p-3-e^2}}
\end{align}
The above expressions are written under slow-ration approximation to the linear orer $\mathcal{O}(a)$. Further, using these constants of motion, one can write down the location of the last stable orbit, i.e., the point where the inspiralling object ends its trajectory, also termed as \textit{separatrix}. Since the actual expression, even upon slow-rotation approximation, is quite lengthy, we provide the separatrix with leading-order corrections in ($a, \ell$) to see how such a deviation appears in the separatrix.
\begin{align}\label{sep}
p_{sp} = 6+2 e-4 \sqrt{2} a \sqrt{\frac{e+1}{e+3}}+\frac{e-3}{e+3}(e+1)\ell +\mathcal{O}(a^2)+\mathcal{O}(\ell^2).
\end{align}
Note that while performing numerical computations in subsequent sections, we use the expression of the separatrix linearized in $a$ with full consideration of $\ell$. The derived expressions comply with \cite{Glampedakis:2002ya, AbhishekChowdhuri:2023gvu} when $\ell\rightarrow 0$ and with \cite{PhysRevD.50.3816, Kumar:2024utz} when ($\ell\rightarrow 0, a\rightarrow 0$).
As mentioned earlier, the motion occurs between ($r_{p}, r_{a}$), we parameterize the orbital motion in the following manner to overcome divergences appearing at the turning points in differential equations,
\begin{align}\label{parame}
r = \frac{p}{1+e\cos\chi},
\end{align}
where $\chi\in (0, 2\pi)$ is an angular parameter that helps compute the rate of change of constants of motion under the radiation reaction over a period. Further, since the eccentric motion possesses two orbital frequencies: azimuthal ($\Omega_{\phi}$) and radial ($\Omega_{r}$), we can write down the frequencies as \cite{AbhishekChowdhuri:2023gvu, Kumar:2024utz, Carson:2020dez, Cutler:1994ys, PhysRevD.93.044010}
\begin{align}\label{frephir}
\Omega_{\phi} = \frac{J_{z}}{Er^{3}}(r-2e^{-\ell/r})+\frac{2ae^{-2\ell/r}}{r^{6}}\Big(e^{\ell/r}r^{3}+\frac{J_{z}^{2}}{E^{2}}(re^{\ell/r}-2) \Big) + \mathcal{O}(a^2)\hspace{3mm} ; \hspace{3mm}
\Omega_{r} = \frac{2\pi}{T_r}
\end{align}
where
\begin{align}
T_{r} =& \frac{\pi}{2 (1-e^2)^{3/2} \sqrt{p}}\Big(-3 (e^2-1) \ell^2 +4 (e^2-1) \ell (\sqrt{1-e^2}+3)+4 p^{2}\Big) \\ \nonumber
& +\frac{2\pi a}{p\sqrt{1-e^2}}\left((2 \sqrt{1-e^2}+9) \ell-3 p\right)+\mathcal{O}(a^2),
\end{align}
$T_r$ denotes the radial time period. We use these expressions, at the later stage of the article, to obtain the fundamental frequencies in terms of eccentric orbital parameters ($p, e$) by taking the average over the time period. Next, we briefly discuss the setup and compute the observable quantities.
\section{Fluxes and orbital evolution} \label{radiation:reaction}

In this section, we derive analytical expressions for the average rate change of orbital energy and angular momentum of the inspiralling object exhibiting eccentric motion in the background of the rotating regular black hole spacetime. As a result, the object's motion generates perturbations in the background spacetime, called GWs. The generation of GWs introduces backreaction on the moving object, giving rise to the notion of radiation reaction. Consequently, the constants of motion no longer remain constant; instead, they evolve over time. We examine our study with the leading-order PN analysis and mass-ratio, and compute the effects of deviation, arising from the regular black hole, to Kerr geometry up to 2PN.

It is useful to introduce Cartesian coordinates in the leading-order PN framework: ($x_1, x_2, x_3$) = ($r\sin\theta\cos\phi, r\sin\theta\sin\phi, r\cos\theta$). We can mention the constants of motion in the following form,
\begin{align}\label{const}
\mathcal{E} = \frac{1}{2}\dot{x}_{i}\dot{x}_{i}-\frac{1}{\sqrt{x_{i}x_{i}}}+\frac{1}{x_{i}x_{i}}(\ell+4) \hspace{3mm} ; \hspace{3mm} J_{z} = \epsilon_{3jk}x_{j}\dot{x}_{k},
\end{align}
where, $\epsilon_{3jk}x_{j}\dot{x}_{k} = r^{2}\sin^{2}\theta\dot{\phi}$. The dot denotes the derivative with respect to the coordinate time. We also set $\theta=\pi/2$ for the equatorial consideration. It is to note that the velocity-dependent terms in Eq. (\ref{const}) will contribute in radiation reaction acceleration while computing the rate of change of energy and angular momentum. Therefore, we ignore velocity-independent terms in Eq. (\ref{const}). With this, the instantaneous rate of change of orbital energy and angular momentum of the inspiralling object due to the radiation reaction can be written as
\begin{align}\label{insflx}
\dot{\mathcal{E}} = x_{i}\Ddot{x}_{i} \hspace{3mm} ; \hspace{3mm} \dot{J}_{z} = \epsilon_{3jk}x_{j}\Ddot{x}_{k}.
\end{align}
The $\Ddot{x}_{i}$ will contribute as a radiation reaction (RR) acceleration, which is often denoted as $a_{j}$. Finally, the radiation reaction acceleration is given by \cite{Flanagan:2007tv, PhysRevD.52.R3159}
\begin{align}\label{accelration}
a_j=-\dfrac{2}{5}I^{(5)}_{jk}x_{k}+\dfrac{16}{45}\epsilon_{jpq}J^{(6)}_{pk}x_{q}x_{k}+\dfrac{32}{45}\epsilon_{jpq}J^{(5)}_{pk}x_{k} \dot{x}_{q}+\dfrac{32}{45}\epsilon_{pq[j}J^{(5)}_{k]p}x_{q} \dot{x}_{k}+\dfrac{8J}{15}J^{(5)}_{3i},
\end{align}
where $I_{jk}$ and $J_{jk}$ are defined as the mass and current quadrupole moments, written as
\begin{equation}\label{moments}
I_{jk}=\Big[x_j x_k\Big]^{\text{STF}} \hspace{3mm} ; \hspace{3mm}
J_{jk}=\Big[x_{j}\epsilon_{kpq}x_{p}\dot{x}_{q}-\dfrac{3}{2}x_jJ\delta_{k3}\Big]^{\text{STF}}.
\end{equation}
The superscripts denote the order of derivative with respect to coordinate time, and `[\hspace{2mm}]' represents the anti-symmetric notations, $B_{ij}=\frac{1}{2}(B_{ij}-B_{ji})$. The notation $`J'$ in the last term of Eq. (\ref{accelration}) refers to the black hole spin $a$. This prescription now enables us to compute the quantity in Eq. (\ref{insflx}). By replacing the parametrization of Eq. (\ref{parame}),
we obtain the instantaneous rate change of constants of motion in terms of ($p, e, \chi$). We further average it over the course of a period and obtain the average rate of change of energy and angular momentum, which we call the average loss of orbital energy and angular momentum over a period. Collectively, it can be written in the following fashion:
\begin{align}
<\Dot{\mathcal{C}}> = \frac{1}{T_{r}}\int_{0}^{2\pi}\Dot{\mathcal{C}}\frac{dt}{d\chi}d\chi \hspace{3mm} ; \hspace{3mm} \mathcal{C}\in (\mathcal{E}, J_{z}),
\end{align}
As a result, for the spacetime under inspection, we arrive at
\begin{equation}
\begin{aligned}\label{flx}
<\Dot{\mathcal{E}}> =& \frac{(1-e^{2})^{3/2}}{15p^{5}}\Big[-\left(37 e^4+292 e^2+96\right)+\frac{24\ell}{p}\left(33 e^4+104 e^2+24\right) + \\
& \frac{a}{2 p^{3/2}}\left(491 e^6+5694 e^4+6584 e^2+1168\right) -\frac{1}{2p^{2}} \Bigl\{ 2\left(1-e^2\right)^{3/2} \left(37 e^4+292 e^2+96\right) \\
& \ell+ \ell \left(e^6 (3779-491 \ell)+5 e^4 (5048-2247 \ell)+4 e^2 (5402-4743 \ell)+96 (26-33 \ell)\right)\Bigl\}\Big]+\mathcal{O}(a^{2}) \\
<\Dot{J_{z}}> =& \frac{4(1-e^{2})^{3/2}}{5p^{7/2}}\Big[-\left(7 e^2+8\right)+\frac{\ell}{p}\left(2 e^4+63 e^2+40\right)+\frac{a}{12 p^{3/2}}\left(549 e^4+1428 e^2+488\right) - \\
& \frac{(1-e^{2})^{-1/2}\ell}{4p^{2}}\Bigl\{(28 e^6-24 e^4-36 e^2-\sqrt{1-e^2} (12 e^6+e^4 (757-140 \ell)+ \\
& e^2 (1636-955 \ell)-360 \ell+352)+32)\Big\}\Big]+\mathcal{O}(a^{2}).
\end{aligned}
\end{equation}
The results comply with existing literature \cite{Zi:2024jla, Kumar:2024utz, Flanagan:2007tv, Glampedakis:2002ya, Ryan:1995xi, PhysRev.131.435, PhysRev.136.B1224}. We have provided expressions up to 2PN order. The cross term $\mathcal{O}(a\ell)$ appears at higher-order beyond 2PN.  One can track the PN orders of distinct terms by restoring the powers of speed of light ($c$). In our case, $<\Dot{\mathcal{E}}>\sim \frac{1}{p^{5}c^{5}}a_{0}\Big(-a_{1}+a_{2}\frac{24\ell}{pc^{2}}+a_{3}\frac{a}{p^{3/2}c^{3}}-a_{4}\frac{1}{p^{2}c^{4}}\Big)$ and  $<\Dot{J}_{z}>\sim \frac{b_{0}}{p^{3}c^{5}}\Big(-\frac{b_{1}}{p^{1/2}}+b_{2}\frac{\ell}{p^{3/2}c^{2}}+b_{3}\frac{a}{p^{2}c^{3}}-b_{4}\frac{1}{p^{5/2}c^{4}}\Big)$. Where ($a_{0}, a_{1}, a_{2}, a_{3}, a_{4}$) and ($b_{0}, b_{1}, b_{2}, b_{3}, b_{4}$) are functions of eccentricity, can be seen in Eq. (\ref{flx}). Note that the leading-order emergence of deviation parameter $\ell$ takes place from the 1PN. It is worth mentioning that one should in principle include the GR PN results up to 2PN, as our focus in the present article is up to 2PN. However, such an inclusion of higher-order GR PN results requires several advancements in the framework \cite{Canizares:2012is}, which may give rise a more stringent bound on the parameter $\ell$. Additionaly, one cannot apriori comment that the flux contribution due to the deviation will be subleading in comparison to the one that comes from GR PN results. Therefore, we only focus on the corrections in observables coming from the non-Kerr/deviation parameter and leave the investigation with GR PN results for future studies containing analysis of 2PN and beyond \cite{Moore:2016qxz, Blanchet:2013haa}.

With the expression (\ref{flx}) in hand, one can now determine the orbital evolution of the inspiralling object. Since we can relate ($\mathcal{E}, J_{z}$)$\longleftrightarrow$ ($p, e$), therefore, in order to obtain the orbital evolution, we use: $\frac{d\mathcal{E}}{dt} = \frac{\partial \mathcal{E}}{\partial p}\frac{dp}{dt} + \frac{\partial\mathcal{E}}{\partial e}\frac{de}{dt}$ and $\frac{dJ_{z}}{dt} = \frac{\partial J_{z}}{\partial p}\frac{dp}{dt} + \frac{\partial J_{z}}{\partial e}\frac{de}{dt}$ which finally gives us the average rate change of ($p, e$). The corresponding expressions take the following form:
\begin{equation}
\begin{aligned}
\Big\langle\frac{dp}{dt}\Big\rangle = \Big(\frac{\dot{\mathcal{E}}\partial_{e}J_{z} -\dot{J}_{z}\partial_{e}\mathcal{E}}{\partial_{p}\mathcal{E}\partial_{e}J_{z}-\partial_{e}\mathcal{E}\partial_{p}J_{z}}\Big) \hspace{3mm} ; \hspace{3mm} \Big\langle\frac{de}{dt}\Big\rangle = \Big(\frac{\dot{J}_{z}\partial_{p}\mathcal{E}-\dot{\mathcal{E}}\partial_{p}J_{z}}{\partial_{p}\mathcal{E}\partial_{e}J_{z}-\partial_{e}\mathcal{E}\partial_{p}J_{z}}\Big)\,.
\end{aligned}
\end{equation}
As a result, we have
\begin{equation}
\begin{aligned}\label{dpdtn}
\Big\langle\frac{dp}{dt}\Big\rangle =&  \frac{2(1-e^{2})^{3/2}}{5p^{3}}\Big[-4\left(7 e^2+8\right) + \frac{8\ell}{p}\left(e^4+35 e^2+24\right)+\frac{a}{3 p^{3/2}}\left(475 e^4+1516 e^2+1064\right)\Big] \\
& +\frac{1}{15p^{5}}\Big[24 (1-e^2)^2 (7 e^4+e^2-8) \ell+(1-e^2)^{3/2} \ell (72 e^6+e^4 (4582-893 \ell) \\
& +e^2 (9544-7376 \ell)-3456 \ell+3264)\Big] +\mathcal{O}(a^{2})\\
\Big\langle\frac{de}{dt}\Big\rangle =& \frac{e(1-e^{2})^{3/2}}{15p^{4}}\Big[-\left(121 e^2+304\right) + \frac{24\ell}{p} \left(e^4+67 e^2+93\right)+\frac{a}{2p^{3/2}} \left(1313 e^4+5592 e^2+7032\right)\Big] \\
& +\frac{1}{30ep^{6}}\Big[-2 e^2 (1-e^2)^3 (121 e^2+304) \ell+(1-e^2)^{3/2} \ell (72 e^8+e^6 (8329-1394 \ell) \\
& +e^4 (27874-17136 \ell)+24 e^2 (686-639 \ell)+384)\Big]+\mathcal{O}(a^{2})
\end{aligned}
\end{equation}
The average rate of change of orbital parameters implies the gradual decrease in ($p(t), e(t)$) in time as long as the contributions from spin $a$ and deviation $\ell$ are subleading. One can again restore the powers of $c$ to count the PN order; however, it goes the same as the case of Eq. (\ref{flx}).

Further, one can also determine how the parameter $\ell$ affects the inspiral timescale. Alternatively, how much time does the inspiralling object take to reach the LSO? Let us consider that the inspiral starts from $p_{in}=14$ and reaches the position near the LSO at $p=p_{sp}$. We integrate Eq. (\ref{dpdtn}) and obtain the timescale by subtracting the GR contributions. This infers the effect of the deviation parameter $\ell$ on the inspiralling timescale. The corresponding expression takes the following form

\begin{align}\label{delt}
\Delta t \approx & \frac{5(1-e^{2})^{-3/2}}{192 (8+7e^{2})^{2}} \Big[-6 \left(7 e^2+8\right) \left(p_{sp}^4-38416\right)-\ell (p_{sp}-14) \Big(36 e^6 (p_{sp}+14)+ \nonumber\\
& 4 e^2 (p_{sp} \left(3 \sqrt{1-e^2}+140 p_{sp}+3153\right)+42 (\sqrt{1-e^2}+1051))-96 (p_{sp} (\sqrt{1-e^2}-4 p_{sp}-73)+ \nonumber\\
& 14 (\sqrt{1-e^2}-73))+e^4 (p_{sp} (84 \sqrt{1-e^2}+16 p_{sp}+2515)+14 (84 \sqrt{1-e^2}+2515))\Big)\Big]+\mathcal{O}(\ell^{2}).
\end{align}
To restate, the $\Delta t$ is obtained by subtracting the GR part of the contribution. Note that here we have linearized the $\Delta t$ in $\ell$ to see the impact of deviation analytically. However, full expression is very huge and not relevant for the present article. If we switch off the parameter $\ell$, the results comply with \cite{Kumar:2024utz}. This calculates how long the test body will need to travel to reach the LSO as a result of the deviation parameter. Next, we investigate the prospects of detectability of the deviation parameter $\ell$ with LISA observations.
\section{Dephasing and waveform}\label{dphasewavefomr}
In this section, we provide the setup for the detectability of the parameter $\ell$ in the context of an EMRI system. The analysis will demonstrate the impact of deviation to the Kerr geometry. As mentioned earlier, the eccentric equatorial orbits will possess two fundamental frequencies ($\Omega_{\phi}, \Omega_{r}$), i.e., azimuthal and radial frequencies. Collectively, we can write down the averaged orbital frequencies in the following way \cite{Rahman:2023sof, AbhishekChowdhuri:2023gvu, Kumar:2024utz, Cutler:1994ys}
\begin{align}
\frac{d\varphi_{i}}{dt} = \langle \Omega_{i}(p(t),e(t))\rangle = \frac{1}{T_{r}}\int_{0}^{2\pi}d\chi\frac{dt}{d\chi} \Omega_{i}(p(t),e(t),\chi) \hspace{5mm} ; \hspace{5mm} i = (\phi, r).
\end{align}
With this, one can derive the analytical expressions of azimuthal frequency \cite{ AbhishekChowdhuri:2023gvu, Kumar:2024utz, Cutler:1994ys, PhysRevD.93.044010, PhysRevD.50.3816}
\begin{align}\label{frephi}
\tilde{\Omega}_{\phi}\equiv\frac{d\varphi_{\phi}}{dt} =& \frac{(1-e^{2})^{3/2}}{p^{3/2}}\Big(1-\frac{\ell}{p}-a\frac{(1+3e^{2})}{p^{3/2}}+\frac{\ell^{2}}{4p^{2}}(1+3e^{2})\ell^{2}\Big) \nonumber \\
& +\frac{\ell(1-e^{2})}{2p^{7/2}}\Big(2+2e^{4}-6\sqrt{1-e^{2}}-e^{2}(4+9\sqrt{1-e^{2}})\Big)+\mathcal{O}(a^{2})
\end{align}
Moreover, the expression for radial frequency is directly readable from Eq. (\ref{frephir}). It is to add that we have written the expression up to 2PN. Further, we can integrate the above equation in order to obtain the orbital phase difference from the Kerr geometry. Such a quantity gives rise the notion of GW dephasing. It can be written as
\begin{align} \label{dephasing}
\Delta \Phi_{\phi,r} = 2\int_0^t \left(\tilde{\Omega}_{\phi,r}^{\ell\neq0}-\tilde{\Omega}_{\phi,r}^{\ell=0}\right) dt
\end{align}
where $\tilde{\Omega}_{\phi,r}^{\ell=0}$ represents the averaged orbital frequency in the Kerr background and $\tilde{\Omega}_{\phi,r}^{\ell\neq0}$ is the averaged orbital frequency for the case of regular black hole. Since we are dealing with quadrupolar radiation here, the GW phase is twice the orbital phase.
\begin{figure}[h!]
\centering
\includegraphics[width=3.17in, height=2.27in]{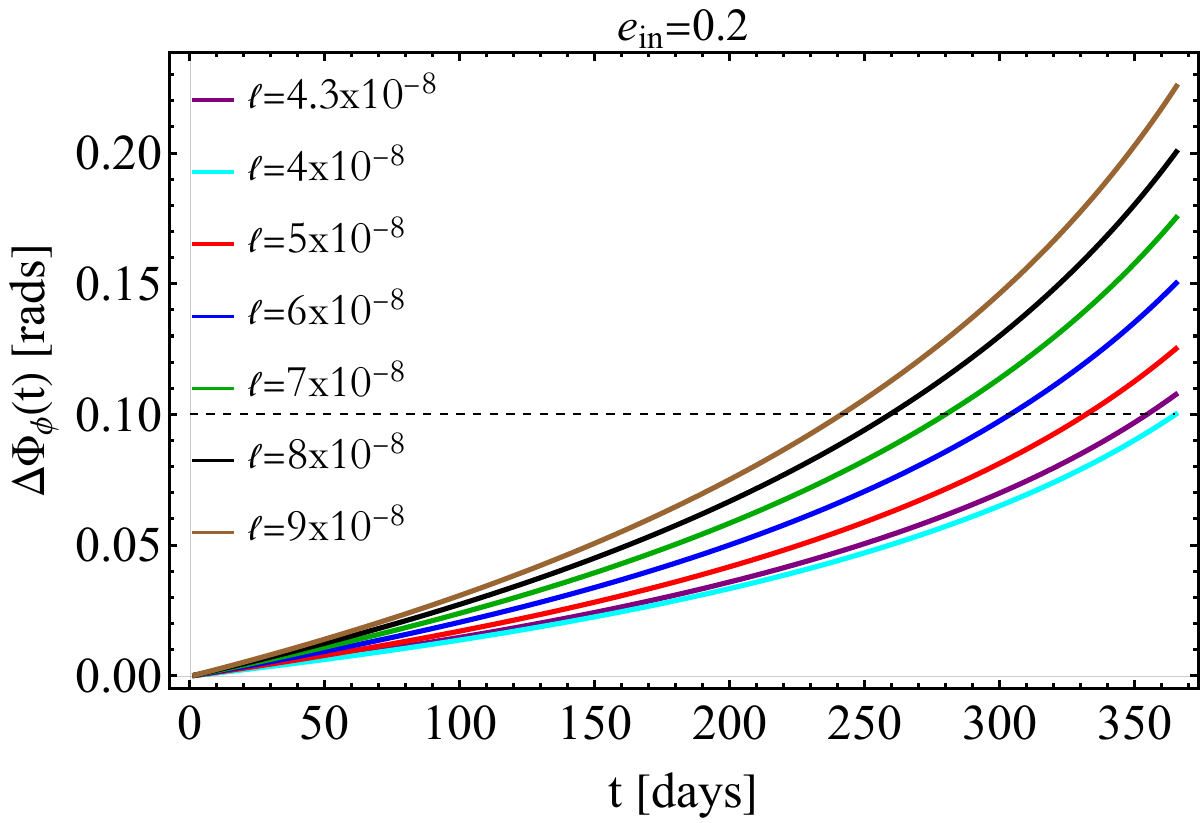}
\includegraphics[width=3.17in, height=2.27in]{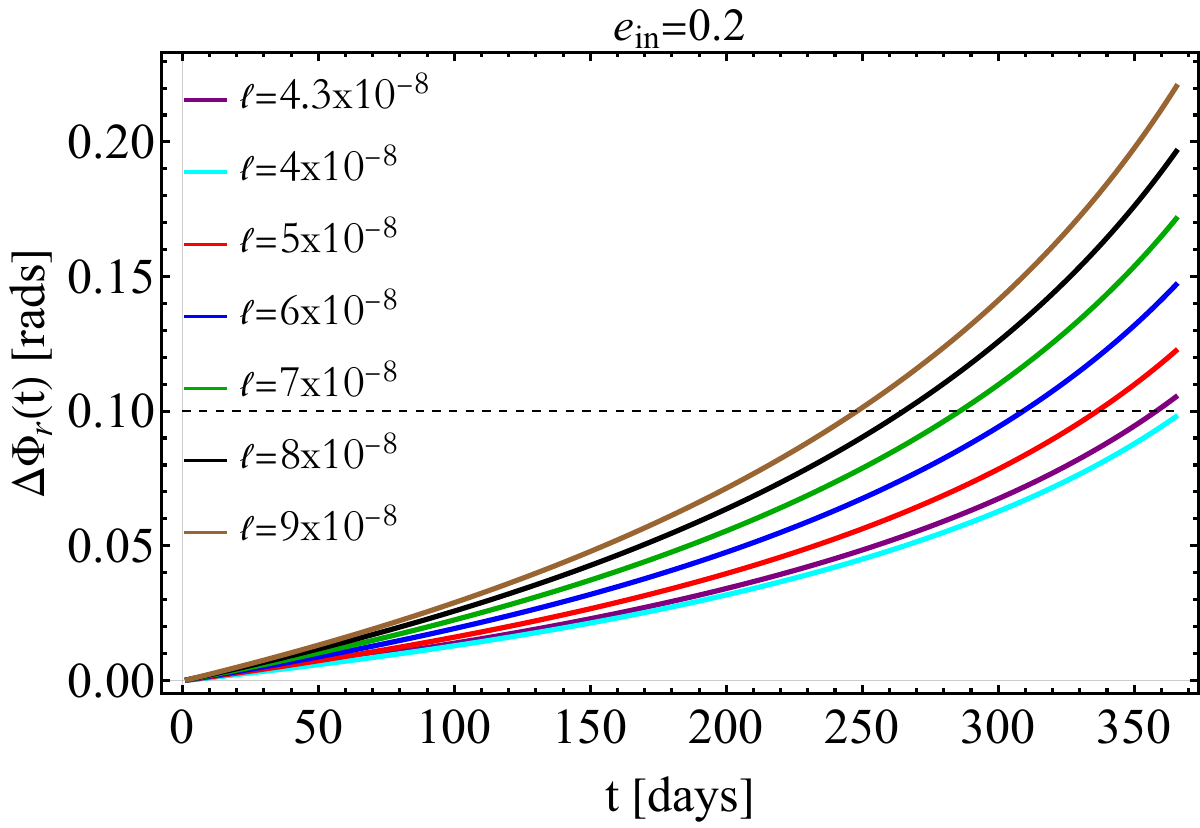}
\caption{The plots show the azimuthal (left) and radial (right) dephasings in one year of the observation period. We have taken mass-ratio $q=10^{-5}$ and the fixed initial eccentricity $e_{\textup{in}}=0.2$, together with distinct values of the parameter $\ell$.} \label{dephasing1}
\end{figure}

We consider the detection criteria of the parameter $\ell$ with the threshold $\Delta\Phi \gtrsim  0.1$ when taking the observation period of one year with the signal-to-noise ratio (SNR) 30 \cite{Bonga:2019ycj,Maselli:2021men}. If we take a closer look at Fig. (\ref{dephasing1}), we notice that the azimuthal dephasing is larger than the radial dephasing. This is consistent with \cite{Zi:2023qfk, Barsanti:2022ana}. Both panels in the above figures depict the observation period for one year. Taking the mass-ratio $q=10^{-5}$ and a fixed initial eccentricity $e_{\textup{in}}=0.2$, we find that for $\ell = 4.3\times 10^{-8}$, the dephasing becomes $ \gtrsim 0.1$, implying that if we go below $\ell\leq4\times 10^{-8}$, LISA might not be able to detect the corresponding dephasing. We also notice when we increase the magnitude of the non-Kerr parameter, the dephasings become larger and more promising from the detection perspective.

\begin{figure}[h!]
\centering
\includegraphics[width=3.17in, height=2.27in]{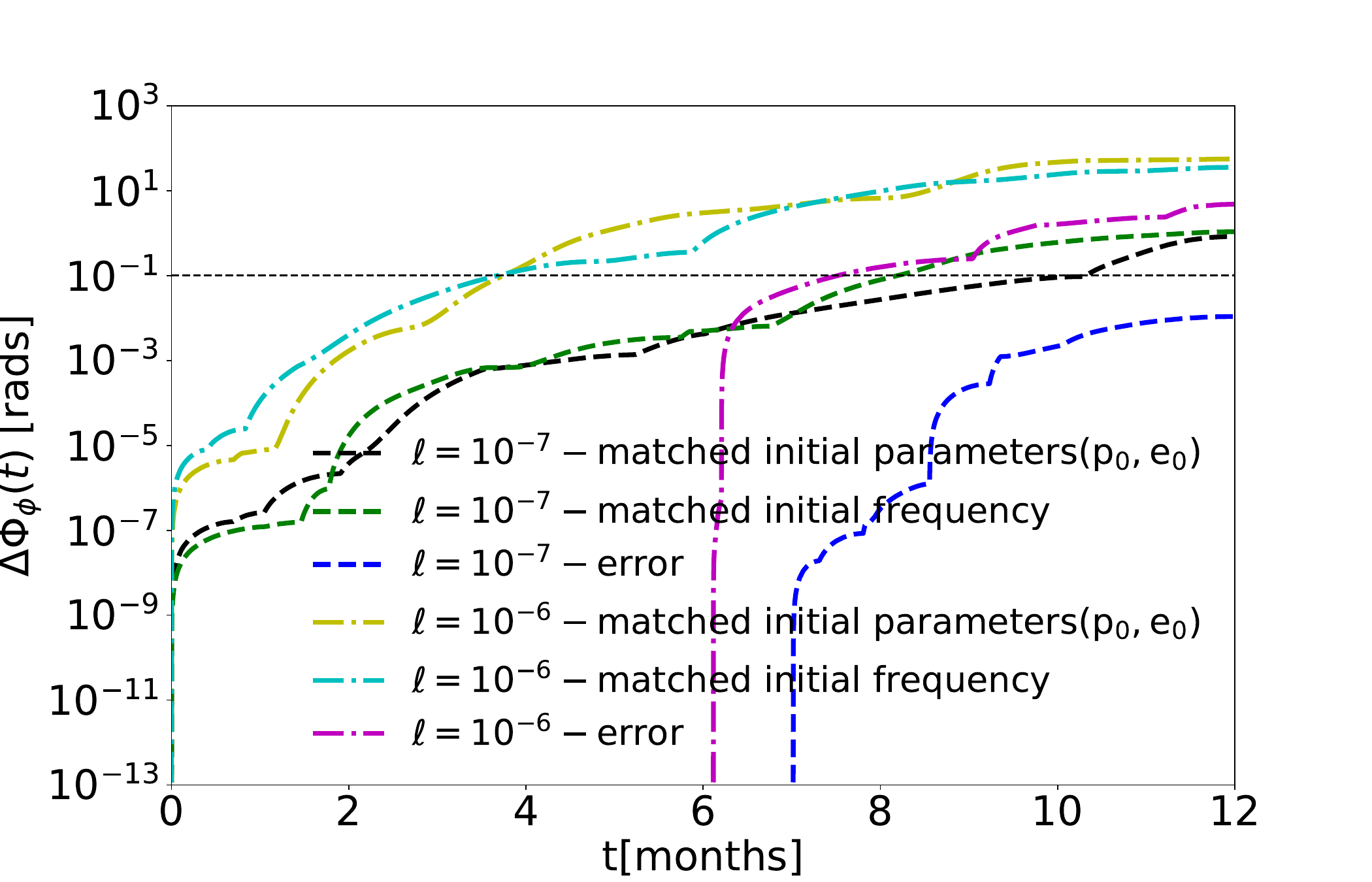}
\includegraphics[width=3.17in, height=2.27in]{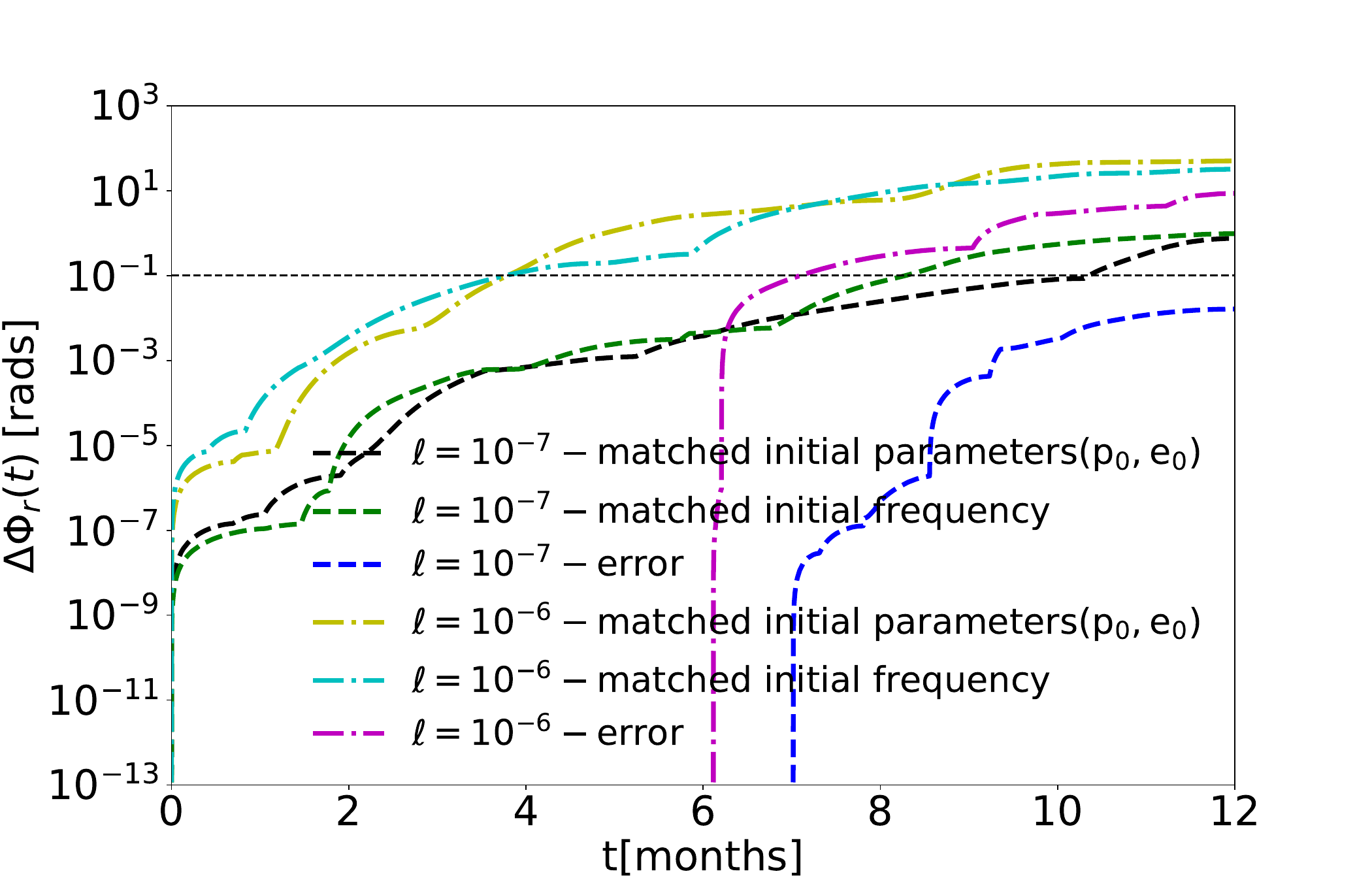}
\caption{Effect of conservative regular parameter $\ell$ corrections on the long time phase evolution is plotted, including the cases of matched initial orbital parameters and frequencies, the label of ``$\rm error$"
the difference of two phase for those above cases. The other parameter keep same with the case of Fig. \ref{dephasing1}.
} \label{dephasing2}
\end{figure}

Given the dephasing of two EMRI waveforms from Kerr and regular rotating black holes, we match the orbital frequencies when evolving the orbital parameters adiabatically, following the method described in Ref.~\cite{Warburton:2011fk}.
To avoid the fake effect of regular rotating BH, the full procedure for evolving EMRI orbital parameters is as follows:
\begin{enumerate}
    \item Set initial parameters \((p_0 = 14.0, e_0 = 0.2)\) and phase \((\Phi_r = 0, \Phi_\phi = 0)\) at \(t = 0\).
    \item Compute the azimuthal and radial orbital frequencies \((\Omega_\phi, \Omega_r)\) at the beginning time.
    \item Evolve the orbital parameters using fluxes given in Eq.~\eqref{flx}, and update the orbital frequencies.
    \item Find the roots \((p, e)\) of the fundamental frequencies equations given in Eqs.~\eqref{frephir} and \eqref{frephi}, the left hand of equations is the frequency values obtained in step 3.
    \item Utilize the parameters $(p,e)$ in step 4 as the initial conditions for the subsequent inspiral iteration.
\end{enumerate}
Repeat steps 2-4 iteratively to evolve the EMRI orbital parameters over time, ensuring the orbital frequencies are consistently matched between the Kerr and regular rotating black hole.
In Fig.~\ref{dephasing2}, we present the azimuthal and radial dephasing as a function of observation time, including the cases of \(\ell = 10^{-7}\) and \(\ell = 10^{-6}\). Note that the two dephasings \(\delta \Phi_{\phi, r}\) are computed considering the matched initial parameters and initial frequencies, the horizontal dashed line is the threshold distinguished by LISA. The label ``$-\rm error$'' denotes the difference between the two phases for the matched initial parameters and frequencies.
One can see that the dephasing grows fast in the beginning and changes slowly over the last few months; after matching the same orbital frequency for two MBH spacetimes, the parameter $\ell$ of regular rotating BH can lead to a distinguishable effect on EMRI orbital phase evolution. It is difficult to judge the dephasing solely from the regular parameter $\ell$ based on Fig. \ref{dephasing1}, as the evolution of the orbital parameters $(p(t), e(t))$ can also contribute to the phase differences even after matched the initial parameters. To cross-check the effect on the GW phase, we refer to Fig. \ref{dephasing2}. However, it should be noted that the dephasing curves in Fig. \ref{dephasing2} appear unsmooth, which may be attributed to errors in finding the roots of the frequency equations in step 4. In fact, obtaining $(p, e)$ requires solving high-order nonlinear binary equations. We utilize the numerical package \textit{Eigen} \cite{eigenweb} within our C++ code to address this problem, where the error in finding roots is approximately $\leq 10^{-2}$.

In the next stage, we perform the waveform calculation and further estimate the bound on $\ell$ through the mismatch calculation together with the estimation of the parameters using FIM. We introduce the approximate waveform formula in the  numerical kludge \cite{Babak:2006uv}, where the smaller perturbation
$h_{ij}(t)$ in the transverse-traceless (TT) can be written as
\begin{equation}\label{eq:inteor}
h_{ij}(t) = \frac{2}{D} \left(P_{ik}P_{jl}-\frac{1}{2}P_{ij}P_{kl}\right)\Ddot{I}^{kl}
\end{equation}
and
\begin{equation}
P_{ij}=\eta_{ij} - \hat{n}_i\hat{n}_j \;.
\end{equation}
Here, the quantity $D$ is the distance from source to detector,
$\eta_{ij}$ is the  flat Minkowski metric, $\hat{n}_i$ is the unit vector of the direction of wave propagation, and $I^{ij}$ is the inertia tensor, which is given by mass quadrupole moment,
\begin{equation}
I^{ij} = \Bigg[\int d^3x x^ix^j T^{tt}(t,x^i)\Bigg]^{\textup{STF}}\;,
\end{equation}
where the superscript $\rm STF$ denotes to the symmetric and trace-free.
For the EMRI binaries, in the point-mass approximation, the non-vanishing components of the stress-energy tensor have $T^{tt}(t,x^i)=\rho(t,x^i)$ and $T^{tj} = \rho(t,x^i) v^j(t)$; $\rho(t,x^i)$ is the energy density of secondaries,
\begin{equation}\label{eq:energydensity:CO}
\rho(t,x^i) = \mu \delta^{(3)}[x^i-z^i(t)]
\end{equation}
where $\delta^{(3)}$ is the three-dimensional Dirac delta distribution, $z^i(t) = dz^i(t)/dt$ is the  spatial Cartesian coordinate, $\mu$ is the mass of the secondary object, and $v^j(t)$ is the spatial velocity.
To compute these stress energy tensor, we can assume the Boyer-Lindquist-like coordinates $(r, \theta, \phi)$ of secondary object as  the flat-space spherical polar coordinate after introducing the  Cartesian coordinate
\begin{equation}
x = r\sin\theta \cos\phi\;, y = r\sin\theta \sin\phi\;, z=r\cos\theta.
\end{equation}
Although there is a non-conservation of the energy-momentum of a particle's motion, leading to inconsistencies with the Boyer-Lindquist coordinate, the approach works well when computing EMRI waveforms in GR \cite{Babak:2006uv}.

Further, the GW polarizations $h_+$ and $h_\times$ can be given by
\begin{equation}
h_+ = \frac{1}{2} \epsilon_+^{ij} h_{ij}(t)  \;,
\quad h_\times = \frac{1}{2} \epsilon_\times ^{ij} h_{ij}(t) ,
\end{equation}
where $\epsilon_{+} ^{ij} = p_ip_j-q_iq_j$ and $\epsilon_{\times} ^{ij} = 2p_{(i}q_{j)}$ are a pair of spatial orthonormal basis; they can be formed by three vectors $(\hat{\mathbf{n}},\hat{\mathbf{p}},\hat{\mathbf{q}})$.
The vector $\hat{\mathbf{n}}$ is the propagation direction of the wave from the EMRI system to the detector on the ecliptic plane; the orthogonal unit vectors $(\hat{\mathbf{p}},\hat{\mathbf{q}})$ are obtained with the vectors ($\hat{\mathbf{n}}$, $\hat{\mathbf{p}}$) and massive black hole (MBH) spinning direction $\mathbf{z}$, $\mathbf{S}= a M \mathbf{z}$, that is $\hat{\mathbf{p}}= \frac{\hat{\mathbf{n}}\times \hat{\mathbf{z}}}{|\hat{\mathbf{n}}\times \hat{\mathbf{z}}|}$ and $\hat{\mathbf{q}}=\hat{\mathbf{p}}\times \hat{\mathbf{n}}$.
From Eqs. \eqref{eq:inteor} and \eqref{eq:energydensity:CO}, the expression of GW polarization can be simplified with the secondary object's vector $v^i=dz^i/dt$ and acceleration $a^i=d^2z^i/dt^2$
\begin{equation}
h_{+,\times} = \frac{2\mu}{D} \epsilon_{+} ^{ij} [a^i(t)z^j(t) + v^i(t)v^j(t)].
\end{equation}
Once the waveforms are generated by the above formula, we can compute the EMRI signal detected by the space-borne GW detector using the response function. The full antenna pattern functions $F_{\rm I,II}(t)$ can be referred to in \cite{Apostolatos:1994mx,Cutler:1994ys,Barack:2003fp}.
\begin{equation}
h_{\rm I,II} (t) = \frac{\sqrt{3}}{2}[F_{\rm I,II}(t) h_+(t) + F_{\rm I,II}(t)h_\times(t) ],
\end{equation}

\begin{figure*}[!h]
\centering
\includegraphics[width=16cm, height =7.4cm]{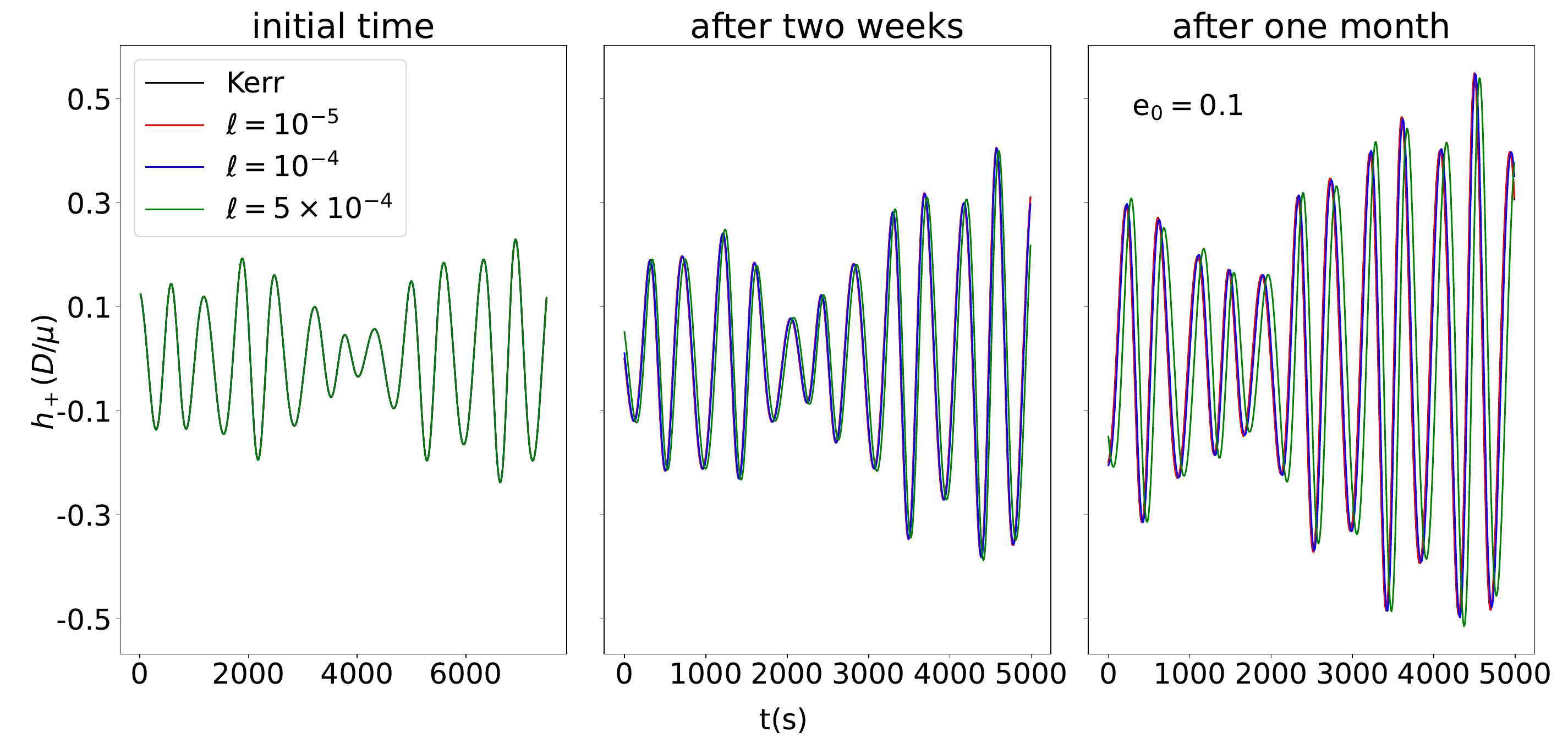}
\includegraphics[width=16cm, height =7.4cm]{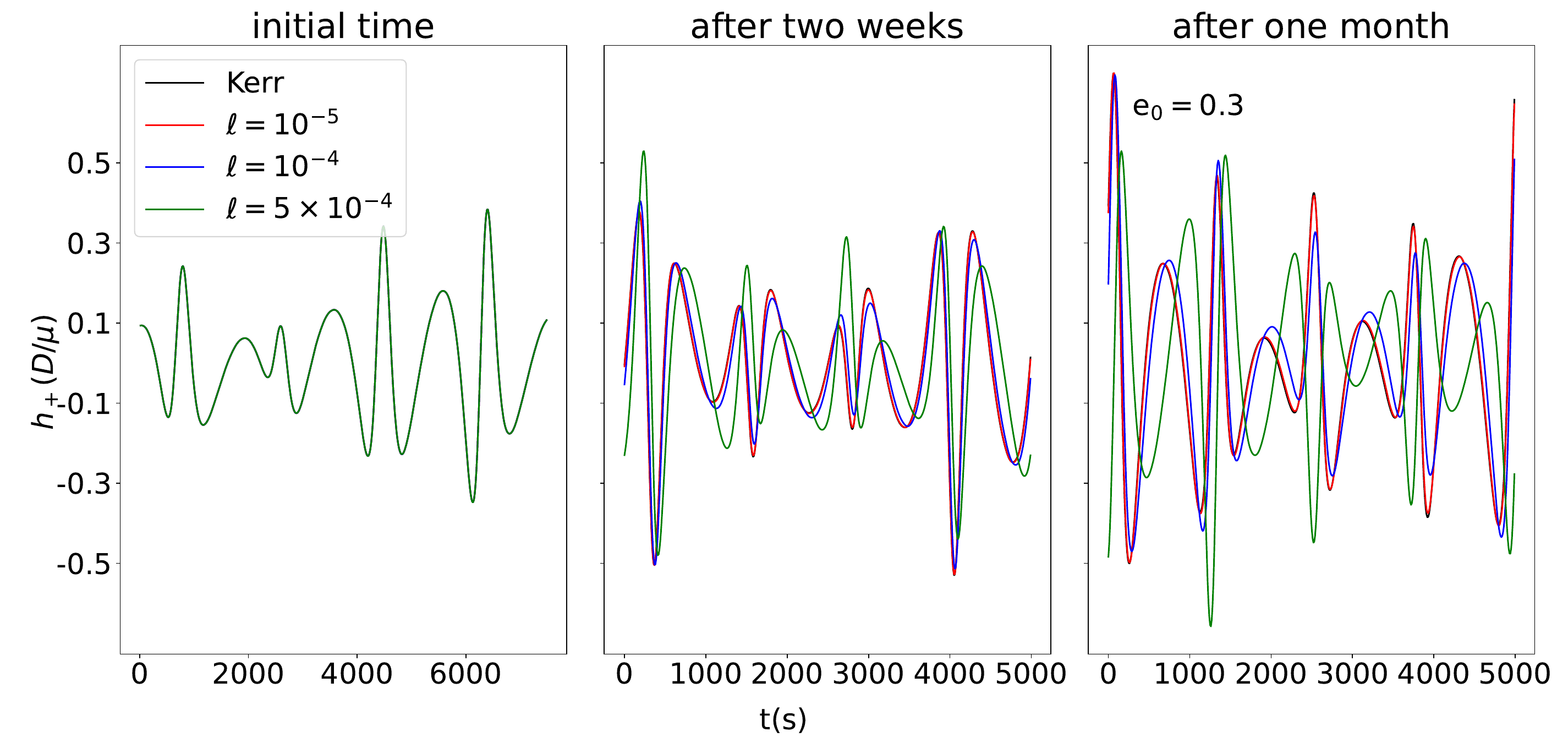}
\caption{Comparison between the polarizations $h_+$ of four EMRI waveforms for the standard Kerr and rotating regular black hole with a parameter $\ell \in\{10^{-5},10^{-4},5\times 10^{-4}\}$ case. The spin of the rotating regular black hole and the initial orbital parameters are set as $a=0.1, p_0=14, e_0\in\{0.1,0.3\}$ and mass-ratio $q=10^{-5}$ and initial phase $\Phi_{\phi,0}=3.0, \Phi_{r,0}=1.0$.
The left panels are the initial stage of the time domain waveforms, the middle panels
are time domain waveforms after orbital evolution of two weeks
and the right panels denote the time domain waveforms after one month.
The three subplots in the top panel show the EMRI waveform setting initial orbital eccentricity $e_0=0.1$, and the bottom panel displays the case of $e_0=0.3$.
}\label{fig2:wave}
\end{figure*}

For the EMRI waveform from the rotating regular MBH, the signal can be determined by fifteen parameters,
\begin{equation}\label{eq:params}
\lambda = \{M,\mu,a,p_0,e_0,\ell,\Phi_{r,0},\Phi_{\phi,0},\phi_s,\theta_s,\phi_K,\theta_K,D\},
\end{equation}
where the quantities with a subscript $0$ represent the initial value of parameters, the angels $(\phi_s,\theta_s)$ denote the direction from the wave source to the solar system barycenter, and the angels $(\phi_K,\theta_K)$ are the direction of the MBH's spin in the  ecliptic-based plane.
To observe the difference in EMRI waveform phase between the Kerr black hole and rotating regular black hole, we plot the comparison of GW
polarizations for two types of black hole spacetimes in Fig. (\ref{fig2:wave}).
It is found that, for several cases of $\ell \in\{0,10^{-5},10^{-4},5\times 10^{-4}\}$, the phases of EMRI waveforms remain the same with each other at the initial $6000$ seconds of orbital evolution for the case of initial orbital eccentricity $e_0=0.1$, there is a slight difference by the orbital evolution of two weeks and there is the difference of phase with the naked eye after about one month. Furthermore, the phase deviation becomes obvious when the orbital eccentricity is larger $(e_0=0.3)$ in the bottom panel of Fig.~\ref{fig2:wave}.
However, it is difficult to quantitatively distinguish the effect of rotating regular black holes from the time domain waveforms; a more rigorous analysis is computing the mismatch of EMRIs waveforms from the spinning regular black hole and Kerr black hole that we perform in the following section.

\section{Detectability}\label{detect}
In this section, we examine the distinguishability and the constraint of the rotating  regular black hole by computing the mismatch and FIM from the EMRIs signal. First, we introduce the method of distinguishing two waveforms by computing the mismatch, which is defined by the inner product,
\begin{equation}
\mathcal{M} = 1- \frac{<h_a|h_b>}{\sqrt{<h_a|h_a><h_b|h_b>}},
\end{equation}
where the noise-weighted inner product $<|>$ can be given by
\begin{equation}
<h_a|h_b> = 2 \int_0^{\infty} \frac{h^\ast_a(f)h_b(f)+h_a(f)h^\ast_b(f)}{S_n(f)}df,
\end{equation}
where the variables $h_{a,b}(f)$ are the Fourier transform of EMRI waveform data in the time domain, the variables $h^\ast_{a,b}(f)$ represents the complex conjugation, and $S_n(f)$ is the noise power spectral density of the GW detector. The rule of thumb to distinguish the two waveforms has been employed widely \cite{Flanagan:1997kp,Lindblom:2008cm}, in which two kinds of GW signals would be identified if their mismatch satisfied the condition $\mathcal{M}\leq 1/(2\rho^2)$. In the realm of the EMRI's source parameter estimation, the threshold of SNR of the signal detected by LISA or TianQin is usually set as $\rho=20$ \cite{Babak:2017tow,Fan:2020zhy}; thus, the minimum value of mismatch distinguished by the detectors is $\mathcal{M}_{\rm min}=0.00125$.

\begin{figure}[h!]
\centering
\includegraphics[width=3.2in, height=2.27in]{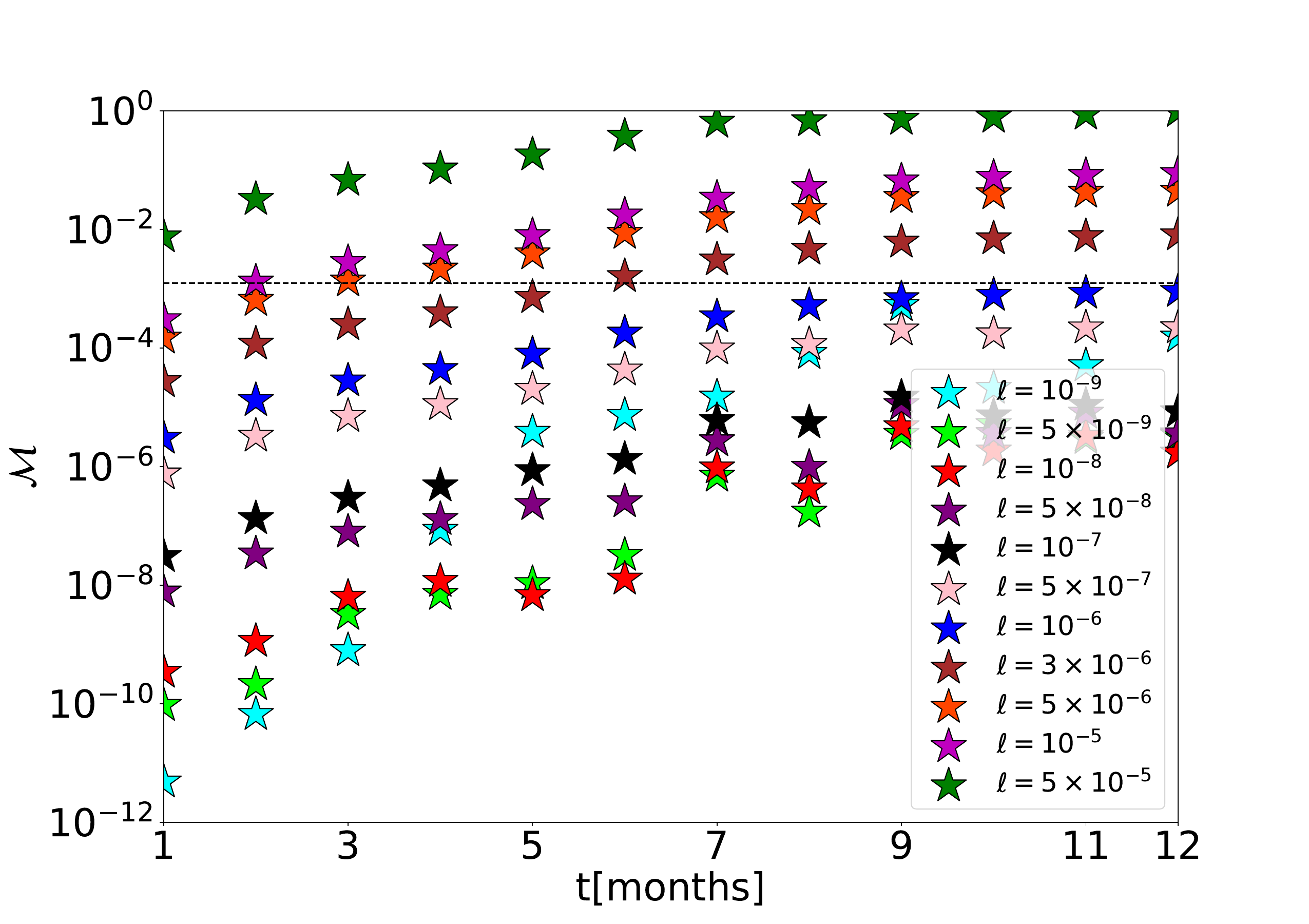}
\includegraphics[width=3.2in, height=2.27in]{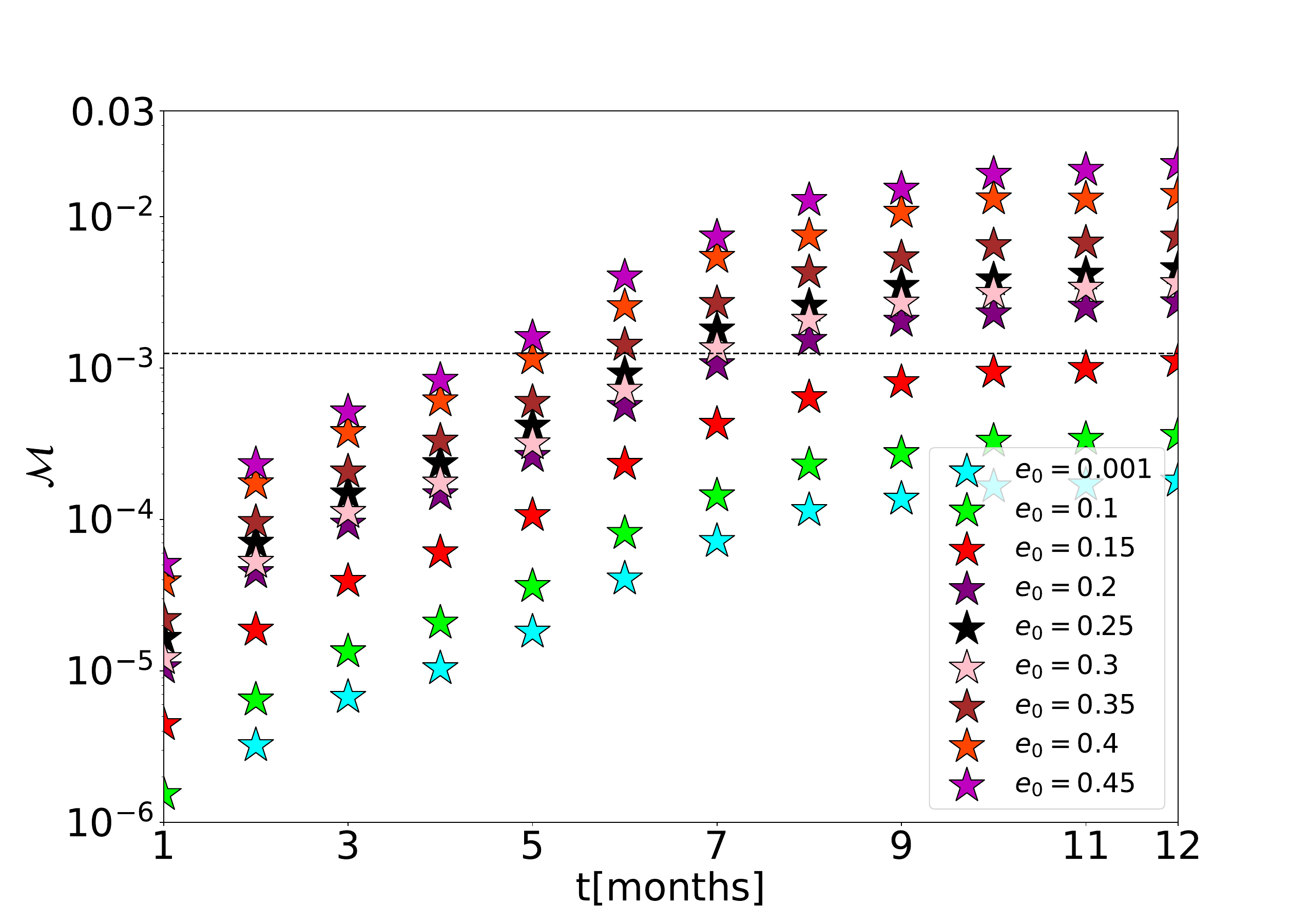}
\caption{The plots show the mismatch as a function of observation time, which considers the mismatch related to a parameter $\ell \in \{10^{-9},5\times10^{-9},10^{-8},5\times 10^{-8},10^{-7},5\times10^{-7},10^{-6},3\times10^{-6},5\times 10^{-6},10^{-5},5\times10^{-5}\}$ in the left panel and an eccentricity $e_0\in \{0.001,0.1,0.15,0.2,0.25,0.3,0.35,0.4,0.45\}$ in the right panel. We have taken mass-ratio $q=10^{-5}$ and the fixed initial eccentricity $e_{0}=0.3$ in the left panel, together with the parameter $\ell = 3\times 10^{-6}$ in the right panel.
The horizontal black dashed lines denote the detection threshold of mismatch discerned by LISA.
} \label{fig3:mismatch:time}
\end{figure}

\begin{figure}[h!]
\centering
\includegraphics[width=3.17in, height=2.27in]{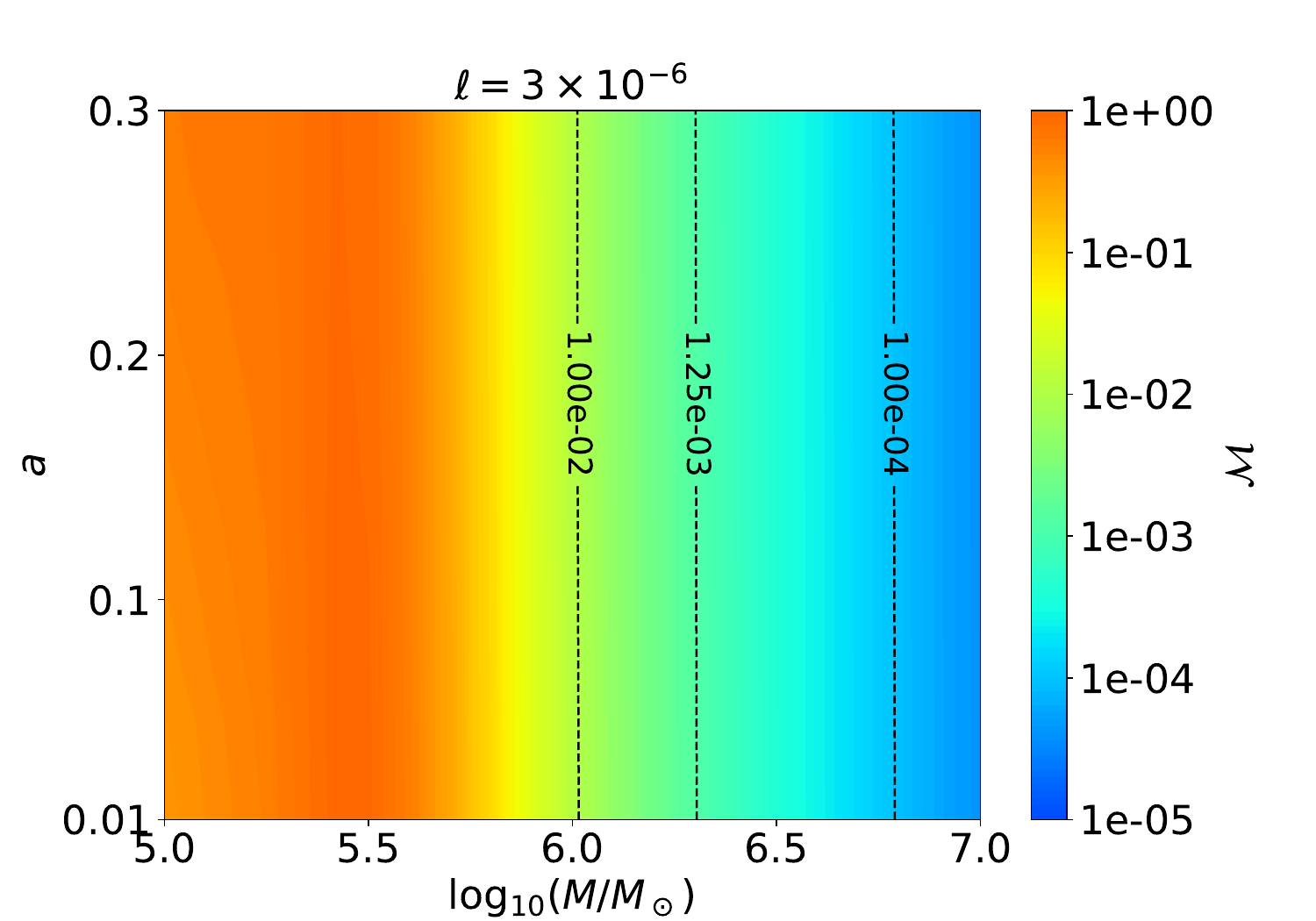}
\includegraphics[width=3.17in, height=2.27in]{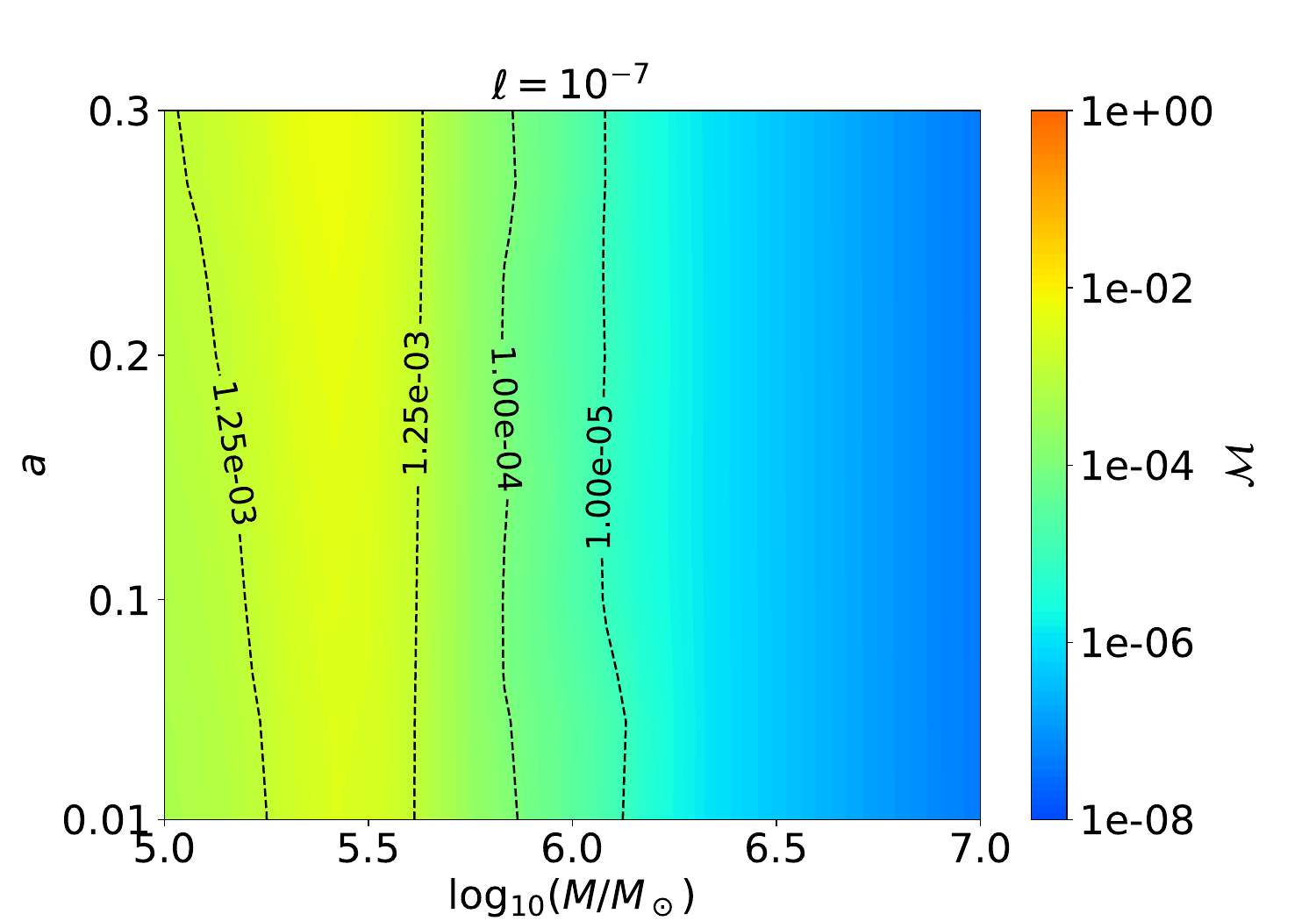}
\caption{Mismatch as a function of mass $\log_{10} (M/M_\odot)$ and spin $a$ of regular MBH is plotted, which includes the cases of $\ell=3\times 10^{-6}$  (left panel) and $\ell= 10^{-7}$ (right panel); the mass-ratio is set $q=10^{-5}$.
The sub-vertical black dashed lines denote the contour value of mismatches,
among which the value $1.25\times 10^{-3}$ of the contour line is just distinguished by LISA.
} \label{fig4:mismatch:contour}
\end{figure}

In Fig. (\ref{fig3:mismatch:time}), the mismatch of two EMRIs waveforms between the Kerr MBH and the rotating regular MBH as a function of observation time is plotted,  which includes the association with the deviation parameter $\ell$ and initial eccentricity $e_0$. We remind that our analysis considers only slow-rotation of the primary with no assumptions on the parameter $\ell$. Note that the horizontal black dashed lines represent the threshold of mismatch distinguished by the detector, and if the mismatch is above the black dashed line, LISA can discern the effect of rotating regular black holes in the EMRIs signal. The left panel, in Fig. (\ref{fig3:mismatch:time}), considers the mismatch for different values of the parameter $\ell$ and a fixed eccentricity $e_0=0.3$, whereas the right panel includes the mismatch for different eccentricities and a fixed value of the parameter $\ell=3\times10^{-6}$. One can see that the mismatch is larger for larger values of the parameter $\ell$; hence, it is obvious that the effect of a regular black hole on the EMRIs waveform is enhanced for the larger parameter $\ell$. From the right panel, we seek to illustrate the relationship between the mismatch and orbital eccentricity. Specifically, the mismatch would become larger when the orbit is more eccentric. Thus, one can claim that the orbital eccentricity plays a key role when considering the
constraint on a regular black hole with the EMRIs signal. The threshold for the parameter $\ell$ discerned by LISA is about $\ell_c \approx 10^{-6}$, which can be bigger when the orbital eccentricity is larger. In order to study the effect of mass and spin of a rotating regular black hole on the mismatch between two EMRIs waveforms, we compute the mismatch as a function of mass $\log_{10}(M/M_\odot)$ and spin $a$ of regular MBH with a parameter $\ell=3\times10^{-6}$ (left panel) and a parameter $\ell=10^{-7}$ (right panel) in Fig. \ref{fig4:mismatch:contour}. The sub-vertical black dashed lines denote the contours of mismatches. LISA can discern the effect of a regular black hole when the mismatch is more than the contour of threshold $\mathcal{M}_c=0.00125$. From the left panel, EMRIs with a mass $M\lesssim 10^{6}M_\odot$ can be distinguished for the case of $\ell=3\times10^{-6}$. Similarly, in the right panel, the EMRIs signal from a regular MBH with a mass of $M\in[10^{-5.25}, 10^{-5.6}]$ can potentially be detected by LISA for the case of $\ell=10^{-7}$. From two panels, one can see that the mismatch depends on MBH mass; it may be due to the effect of the linear appoximation of the MBH spin ($a$) in average losses of energy and angular momentum as well as in orbital evolution, possibly implying the dominance of the mass of the MBH over the spin and making the mismatch less sensitive to the spin of the MBH. This highlights the need for future studies to relax the spin approximation of the MBH.

Next, to perform parameter estimation using the EMRI signal, we employ the FIM method to evaluate measurement errors of the source parameters. Specifically, the analysis is carried out around EMRI source parameters using the analytic hybrid waveform and the noise power spectrum density of LISA~\cite{amaroseoane2017laserinterferometerspaceantenna}. The inverse Fisher matrix yields the parameter covariance: its diagonal entries correspond to the $1\sigma$ uncertainties, whereas the off-diagonal entries capture correlation. This covariance defines a multivariate Gaussian distribution ~\cite{Cutler:1994ys,Vallisneri:2007ev}, which is visualised in the corner plot. The resulting constraints represent forecasted uncertainties under the Gaussian approximation, rather than posteriors derived from noisy mock data injections.

To start with the procedure, in the context of FIM, if the signal-noise-ratio (SNR) $\rho$ is large enough, the covariance of the posterior probability distribution gives the expectation value of the errors $\Delta \lambda^i$,
\begin{align}
<\delta \lambda^i \delta \lambda^j> = (\Gamma^{-1})^{ij} + \mathcal{O}(1/\rho)^{-1} \simeq \Sigma^{ij},
\end{align}
where $\Gamma^{ij}$ is the FIM, given as
\begin{equation} \label{FIM:gamma}
\Gamma^{ij} = <\frac{\partial \bf{h}}{\partial \bf{\lambda}^i}|\frac{\partial \bf{h}}{\partial \bf{\lambda}^j}>.
\end{equation}
So one can approximately compute the measurement error of a parameter by the inverse matrix;
\begin{equation}
\Delta\lambda^i = \sqrt{< (\delta \lambda^i) >^2} \simeq \sqrt{(\Gamma^{-1})^{ii}},
\end{equation}
The detailed discussion can be found in Ref. \cite{Vallisneri:2007ev}.
Additionally, the SNR can be defined with the inner product $<|>$
associated with the power spectral density $S_n(f)$
\begin{equation}
\rho = \sqrt{<\bf{h}|\bf{h}>}=2\sqrt{\int_{f_{low}}^{f_{up}} \frac{\tilde{h}(f) \tilde{h}^\ast(f) }{S_n(f)} df},
\end{equation}
where $f_{low} = 0.1 \rm m Hz$, $f_{up} = f_{\rm LSO}$ is the orbital frequency of the LSO, and the full expression of $S_n(f)$ is found in Append \ref{appC}. The quantity $\tilde{h}(f)$ denotes the Fourier transform of the time-domain signal $h(t)$, and its complex conjugate is $\tilde{h}^\ast(f)$.
\begin{table}[h!]
\centering{\begin{tabular}{lccc}
\hline\hline
$a$          & $0.1$ & $0.1$ & $0.1$
  \\
$e_0$      & $0.1$ & $0.3$ & $0.5$
  \\
\hline\hline
$\Delta(\ln  M)$   & $1.21\times 10^{-4}$   & $7.62\times 10^{-5}$   & $9.96\times 10 ^{-6}$
                 \\ \hline
$\Delta (\ln \mu)$
& $6.83\times 10^{-5}$   & $7.52\times 10^{-5}$    & $4.05\times 10^{-6}$
                 \\ \hline
$\Delta(a)$       & $3.23\times 10^{-6}$   & $3.23\times 10^{-6}$     & $1.02\times 10^{-7}$
                  \\ \hline
$\Delta (p_{0})$   & $1.52\times 10^{-5}$   & $1.43\times 10^{-5}$    & $1.67\times 10^{-7}$
                 \\ \hline
$\Delta (e_0)$     & $4.05\times 10^{-6}$   & $4.0\times 10^{-7}$    & $6.74\times 10^{-7}$
                  \\ \hline
$\Delta (\ell)$     & $1.5\times 10^{-5}$   & $1.02\times 10^{-5}$   & $4.61\times 10^{-7}$
                  \\ \hline
$\Delta(\Phi_{\phi,0})$   & $1.54$    & $2.41$    & $1.08$
    \\ \hline
$\Delta(\Phi_{r,0})$    & $2.41\times 10^{-4}$  & $2.57\times 10^{-6}$   & $5.64\times 10^{-7}$
                 \\ \hline
$\Delta(\Omega_S)$  & $8.25\times 10^{-3}$  & $5.41\times 10^{-3}$   & $2.41\times 10^{-4}$
                 \\ \hline
$\Delta (\Omega_K)$  & $1.25$    & $1.55$    & $1.34$
                 \\ \hline
$\Delta(\ln D)$    & $2.74\times 10^{-1}$  & $7.73\times 10^{-2}$   & $1.24\times 10^{-2}$
 \\
\hline\hline
\end{tabular}}
\caption{\protect\footnotesize
Measurement errors of EMRIs parameters for the inspiral of a $10 M_\odot$
CO onto a $10^6 M_\odot$ MBH at SNR $\rho=20$.
Shown are results for various values of the initial eccentricity $e_0$  and regular MBH spin $a=0.1$.
The other parameters are set as follows:
$\ell=10^{-5}$,
$\Phi_{\phi,0}=1.0$,
$\theta_S=\pi/4$,
$\phi_S=0$,
$\Phi_{r,0}=1.0$,
$\theta_K=\pi/8$,
$\phi_K=0$.
}
\label{table:fims}
\end{table}

\begin{figure*}[ht]
\centering
\includegraphics[width=14.7cm, height =12.3cm]{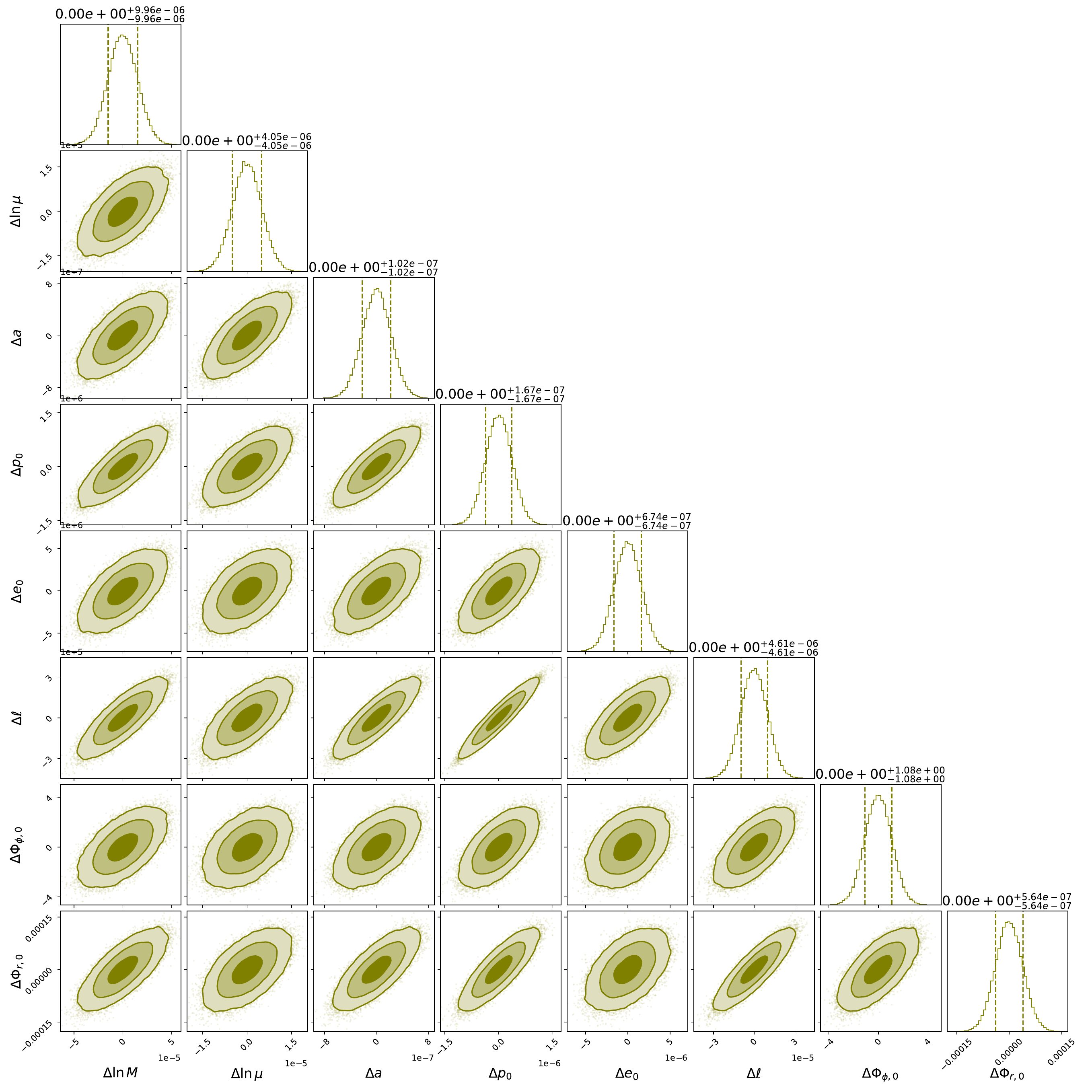}
\caption{Corner plot of the probability distribution for the mass of secondary object, mass and spin of regular MBH, initial orbital eccentricity, semi-latus rectum, initial radial and azimuthal angle variables $(\ln \mu, \ln M,  a=0.1, e_0=0.5, p_0=14,  \ell=10^{-4},\Phi_{\phi,0}=3.0, \Phi_{r,0}=1.0)$ is inferred rom one year observation of EMRIs. The mass-ratio of binary objects is set to $q=10^{-5}$.
The vertical lines denote $1~\sigma$ interval for every source parameter. Three contours show the probability confidence intervals of  $68\%$,  $95\%$ and $99\%$.}
\label{fig:conerplot}
\end{figure*}

Finally, we present the constraint on EMRIs parameters \eqref{eq:params} using the FIM method, which can be listed in Table \ref{table:fims}, where the rotating regular MBH is set as $a=0.1$ and the semi-latus rectum as $p_0=14$. Note that we define solid angles $\Omega_{S,K}$ to succinctly show the measurement error of directional angels  \cite{Babak:2017tow}, and the errors of solid angles can be given by $\Delta\Omega_{S,K} = 2\pi\vert\sin\theta_{S,K}\vert \sqrt{\Sigma^{2}_{\theta_{S,K}}\Sigma^{2}_{\phi_{S,K}}-\Sigma^{2}_{\theta_{S,K}\phi_{S,K}}}$. From the constraint results in Table \ref{table:fims}, one can find that each parameter can be bounded significantly when the orbital eccentricity is bigger; the constraint on the parameter $\ell$ is about the fractional error of $10^{-5}$ using the one-year observation of LISA.
In the following section, we study the correlations for measurement error of
EMRIs parameters under investigation. Given that the orbital frequencies are heavily subjected to the intrinsic parameters, the data analysis and parameter estimation in the EMRI detection mainly pay attention to the correlation of the intrinsic parameters \cite{Babak:2009ua,Wang:2012xh,Chua:2021aah,Katz:2021yft,Ye:2023lok,Zhang:2023vok}. Therefore, we show the probability distribution between the parameter $\ell$ and other parameters $(\ln M,\ln \mu, a, p_0, e_0, \Phi_{\phi,0}, \Phi_{r,0})$.
To assess the relevance among parameters of GW sources, following the simliar method~\cite{Zhang:2022hbt,Zi:2023geb,Zi:2024lmt}, we carry out an analysis of correlation of intrinsic parameters using the off-diagonal elements defined by Eq. \eqref{FIM:gamma}.
According to the corner plot in Fig. (\ref{fig:conerplot}), there is a strong positive
correlation between parameter $\ell$ and other intrinsic parameters $(\ln M, \ln\mu,a,p_0,e_0)$. From the FIM analysis, the prospects of detecting the potential signature of the parameter $\ell$ with LISA-like detectors rely on obtaining stringent constraints on the other intrinsic parameters $(\ln M, \ln \mu, a, p_0, e_0)$.
When computing the inner product of FIM, we adopt a numerical differential interval to evaluate the derivative of the waveform with respect to the parameters $\lambda$. The artificial choice of differential interval may have an influence on the stability of FIM, so we analyze the stability of FIM, presented in Appendix (\ref{fim}).

\section{Discussion}\label{discussion}
EMRIs are one of the promising candidates for probing numerous
characteristics of black holes, including their strong and weak gravity regimes. The detection prospects of such sources with future space-based detectors will bring the foundational changes and deep understanding of gravitational physics. Since black holes are expected to be rotating in general with possible finite eccentricity, it is essential to model these scenarios from detection perspectives. On the other hand, we know that black holes in nature can also differ from the black holes that appear as vacuum solutions of Einstein's theory of GR, such as Schwarzschild and Kerr. Any deviation from these will lead to non-Kerr or Kerr-like geometry in general. These spacetimes not only replace the singular region of the existing black holes by becoming regular but also setup a stage for the observational community, along with some physical constraints such as axisymmetric, asymptotic flatness, separability of the Hamilton-Jacobi equations, energy condition, etc. This spacetime captures an additional parameter, a non-Kerr parameter, whose observational consequences will put forward notable contributions in the line of testing the no-hair theorem and non-GR aspects of the physics.

To restate, we consider an EMRI system where the inspiralling object moves in the vicinity of a central supermassive rotating regular black hole. We adopt the slow-rotation approximation to derive the results, i.e., linearized in spin $a$. With this, we take eccentric equatorial motion of the secondary object and compute analytical expressions for the rate of change of orbital energy and angular momentum up to 2PN and under the effect of radiation reaction. The inclusion of the respective GR PN corrections to the rate of change of orbital energy and angular momentum, including other related quantities, requires several advancements in the current setup, which we leave for the future study \cite{Canizares:2012is}. We further compute GW dephasing and waveform for distinct values of the deviation parameter, indicating notable differences from the Kerr geometry. We further estimate the bound on the non-Kerr parameter with FIM computation using LISA observations. The deviation parameter distinguished by LISA, through mismatch computation, can be reached $\sim 10^{-6}$ with a one-year observation period. The measurement error of the non-Kerr parameter $\ell$ can be confined within a fraction of $\sim 10^{-5}$, depending on the initial orbital eccentricity. In the end, we also analyze the correlation among the non-Kerr parameter $\ell$ and the other intrinsic parameters; as a result, we find that the constraint on the parameter $\ell$ is related to the parameters $(\ln M, \ln \mu, a, p_0, e_0, \Phi_{\phi,0}, \Phi_{r,0})$.

As our present analysis focuses on the equatorial eccentric EMRIs system under the weak-field and slow-spinning approximation. The future works should consider the evolution of general inspiral orbits (non-equatorial) with GW fluxes up to 2PN and beyond \cite{Moore:2016qxz, Blanchet:2013haa, Zi:2025lio} containing the respective general relativistic PN corrections, as well as the relativistic EMRIs fluxes using perturbation theory \cite{Hughes:2021exa, Brito:2023pyl,Khalvati:2024tzz}. As our study explores the order of magnitude anaylsis of the non-Kerr parameter with the potential detectability from LISA observations, it is worth mentioning that the response of the EMRIs signal is based on the condition of low frequency approximation; therefore, the future true EMRIs detection depends on the time-delay interferometry technology \cite{Tinto:2020fcc,Tinto:2023ouy,Wang:2021owg,Tinto:2022zmf, Wang:2021jsv, Wang:2022nea, Wu:2023key}. So, the constraint result needs to be updated through advanced frameworks and the upcoming time-delay interferometry in the near future. Additionally, the constraint on regular MBH is computed with the FIM method; it would be interesting to infer more rigorous results from the Bayesian Markov Chain Monte Carlo-based method \cite{Chua:2021aah,Katz:2021yft}. Furthermore, it will be interesting to explore the regular black holes arising in pure gravity, where solutions are modifications of the standard Schwarzschild black hole and come from a family of gravitational theories called quasitopological gravity, adding higher-curvature elements to the Einstein-Hilbert action \cite{Bueno:2024dgm}. We aim to deliver some of the mentioned problems in our future studies.

\section*{Acknowledgments}
The research of S. K. is funded by the National Post-Doctoral Fellowship (N-PDF: PDF/2023/000369) from the SERB-ANRF, Department of Science and Technology (DST), Government of India.
T.Z. is funded by the China Postdoctoral Science Foundation with Grant No. 2023M731137 and the National Natural Science Foundation of China with
Grant No. 12347140 and No. 12405059.

\appendix

\section{Geodesic equation and constants of motion} \label{apenteu1}
This section provides an overview of the geodesic velocities of the inspiralling object moving in the background of Eq. (\ref{metric}). We use these velocities for the computation of orbital energy and angular momentum loss, phase, and other related quantities. It is to add that we restrict the motion of the inspiralling object on the equatorial plane ($\theta = \pi/2$). This implies the Carter constant ($\mathcal{Q}=0$). However, we write down the velocities in a generic way and set equatorial consideration at a later stage. One can obtain two velocities corresponding to two conserved quantities, energy and angular momentum. For rest, we implement Hamilton-Jacobi method to obtain separable radial and angular velocities \cite{Kumar:2024utz, Yagi:2023eap}
\begin{align}\label{ac1}
S = -\frac{1}{2}\mu^{2}\tau-Et+J_{z}\phi+R(r)+\Theta(\theta) \hspace{3mm} ; \hspace{3mm} -\frac{\partial S}{\partial\tau} = \frac{1}{2}g^{\mu\nu}\frac{\partial S}{\partial x^{\mu}}\frac{\partial S}{\partial x^{\nu}}.
\end{align}
As a result, following \cite{Kumar:2024utz, AbhishekChowdhuri:2023gvu, Zi:2024jla}, we get
\begin{equation}
\begin{aligned}\label{j1}
\mu\frac{dt}{d\tau} =& \frac{1}{\Delta\Sigma}\Big[\left(a^2+r^2\right) \left(a^2 E-a J_{z}+E r^2\right)+a \Delta \left(J_{z}-a E \sin ^2(\theta )\right) \Big] \\
\mu\frac{d\phi}{d\tau} =& \frac{1}{\Delta\Sigma}\Big[a \left(a^2 E-a J_{z}+E r^2-E \Delta\right)+J_{z} \csc ^2(\theta ) \Delta\Big] \\
\mu^{2}\Sigma^{2}\Big(\frac{dr}{d\tau}\Big)^{2} =& \Big(E(a^{2}+r^{2})-aJ_{z}\Big)^{2}-\Delta(\kappa+\mu^{2}r^{2}) \\
\mu^{2}\Sigma^{2}\Big(\frac{d\theta}{d\tau}\Big)^{2} =& (\kappa - \mu^{2}a^{2}\cos^{2}\theta)-\Big(aE\sin\theta-\frac{J_{z}}{\sin\theta}\Big)^{2},
\end{aligned}
\end{equation}
where  $\Sigma = r^{2}+a^{2}\cos^{2}\theta$ and $\Delta = r^{2}+a^{2}-2M r e^{-\ell/r}$. We introduce a more conventional Carter constant ($\mathcal{Q}$) in terms of the separability constant $\kappa$, i.e., $\mathcal{Q}\equiv \kappa-(J_{z}-aE)^{2}$. With this, one can write down the following quantities:
\begin{align}
\mu^{2}\Big[\Big(\frac{dr}{dt}\Big)^{2}+r^{2}\Big(\frac{d\theta}{dt}\Big)^{2}+r^{2}\sin^{2}\theta\Big(\frac{d\phi}{dt}\Big)^{2} \Big] =& \frac{\mu ^2 \left(E^2- \mu ^2\right)}{E^2} +\frac{2\mu ^2 M \left(3 \mu ^2-2 E^2\right)}{E^2 r} \nonumber \\
& +\frac{\mu^{2}}{E^{2}r^{2}}\Big(E^{2}(r-2M)^{2}-(r^{2}-6Mr+12M^{2})\mu^{2}\Big).
\end{align}
Note that we have ignored terms with $\mathcal{O}(\frac{a M}{r^{3}})$ and beyond. Undoubtedly, one can do the higher order expansion in $r$, however, that provides sub-leading results in the leading order PN analysis. Further, to separate the rest-mass energy, we make use of $E=\mu+\mathcal{E}$ \cite{Misner:1973prb, Ryan:1995xi, Flanagan:2007tv}, and discard terms $(\frac{\mathcal{E}M}{r})$ together with their higher-order powers. We finally arrive at
\begin{align}\label{ener2}
\mathcal{E} =& \frac{\mu}{2}\Big[\Big(\frac{dr}{dt}\Big)^{2}+r^{2}\Big(\frac{d\theta}{dt}\Big)^{2}+r^{2}\sin^{2}\theta\Big(\frac{d\phi}{dt}\Big)^{2} \Big]-\frac{\mu M}{r}+\frac{\mu M}{r^{2}}(\ell+4M).
\end{align}
It is to be noted that the velocity-independent terms in Eq. (\ref{ener2}) will not contribute to the computation of energy loss, as the instantaneous losses include the acceleration terms given by Eq. (\ref{insflx}). Further, we can write down following expressions of the velocities from Eq. (\ref{j1})
\begin{equation}
\begin{aligned}\label{gdscs3n}
\mu^{2}\Big(\frac{dr}{d\tau}\Big)^{2} =& 2\mathcal{E}-\frac{2}{r}-\frac{2\ell}{r^{2}}-\frac{J_{z}^{2}}{r^{2}}-\frac{4aJ_{z}}{r^{3}}+\frac{\ell^{3}}{r^{3}}+\frac{4aJ_{z}\ell}{r^{4}}-\frac{2J_{z}^{2}\ell}{r^{4}} \\
\mu\frac{d\phi}{d\tau} =& \frac{J_{z}}{r^{2}}+\frac{2a}{r^{3}}-\frac{2a\ell}{r^{4}}-\frac{8a\ell}{r^{5}}+\frac{a\ell^{2}}{r^{5}}.
\end{aligned}
\end{equation}
Note that we have written Eq. (\ref{gdscs3n}) in dimensionless unit which has been directly used in deriving results in the main text as mentioned in section (\ref{BHspacetime}) and onwards. We have also considered the slowly spinning scenario of the central supermassive black hole which enables us to have analytical expressions of the observables. We are not approximating the parameter $\ell$. The only assumption is set on $a$ with the $r$ expansion about infinity as we are examining the leading-order PN analysis.

Using angular part of the velocity from Eq. (\ref{j1}), we can write down the following expression:
\begin{align}\label{dphidt}
J_{z} = \mu r^{2}\sin^2\theta\frac{d\phi}{d\tau}-\frac{2 a E M \sin^2\theta }{r^{2}}(2 M+(1+\ell)r).
\end{align}
Again, the velocity-independent terms will no longer affect the angular momentum loss of the inspiralling object. The reason goes the same as mentioned before: the instantaneous losses will require acceleration, which will arise from the velocity-dependent terms. Therefore, only the first term will contribute to the computation of Eq. (\ref{insflx}). The derived results are consistent with \cite{Flanagan:2007tv, Ryan:1995xi, AbhishekChowdhuri:2023gvu, Kumar:2024utz, Zi:2024jla} when $\ell\rightarrow 0$. 

\section{Stability of Fisher information matrix}\label{fim}
In this section, we attempt to assess the stability of the perturbed Fisher matrices following the previous works \cite{Piovano:2021iwv,Zi:2023pvl,Speri:2021psr}.
The full technological process can be summarized as follows:
first we construct a matrix $\bf{R}$ with the same dimensionality of Fisher matrix
$\bf{\Gamma}$, whose elements are a uniform distribution $U\in[u_a,u_b]$; then we calculate
the inverse $(\bf{R}+\Gamma)^{-1}$ and the maximum value of relative error with respect to the Fisher matrix,
\begin{equation}
\delta_{\rm stability} \equiv \mathbf{max}_{\rm ij}
\left[\frac{((\bf{\Gamma}+\bf{R})^{-1} - \Gamma^{-1})^{ij}}{(\bf{\Gamma}^{-1})^{ij}} \right],
\end{equation}
where $\bf{R}$ is also named the deviation matrix.
To study the stability of FIM using the EMRIs signal from the regular MBH with a spin
$a=0.1$, we list the stability results $\delta_{\rm stability}$ for different source parameter configurations in Table \ref{FIMstability}.

\begin{table*}[!htbp]
\caption{The stability $\delta_{\rm stability}$ of FIM using the EMRIs signal from the regular MBH with mass $M=10^{6}M_\odot$, spin $a=0.1$ and $\ell=10^{-5}$ is listed, considering the different initial orbital eccentricities.}\label{FIMstability}
	\begin{center}
		\setlength{\tabcolsep}{5mm}
		\begin{tabular}{|c|c|c|c|}
			\hline
			\multirow{2}{*}{$U$}& \multicolumn{3}{|c|}
			{initial orbital eccentricity  $e_0$}
            \\
           \cline{2-4}
			& $0.1$ &$0.3$ &$0.5$ \\
            \hline
		$\in[-10^{-7},10^{-7}]$
            & $4.46\times10^{-1}~ ~$ &$6.57\times10^{-2}$  &$3.04\times10^{-2}$		
              \\		
		\hline
		$\in[-10^{-9},10^{-9}]$
            & $5.57\times10^{-2}~~$ &$4.37\times10^{-2}$ &$1.56\times10^{-2}$
		  \\
	    \hline	
		\end{tabular}
	\end{center}
\end{table*}

\section{Sensitivity curve of LISA}\label{appC}
For the space-borne GW detector LISA, the sky-averaged detector sensitivity can be given by  \cite{LISA:2017pwj,Babak:2017tow}
\begin{eqnarray}
S_n(f)&=&\frac{20}{3}\frac{4S_{n}^\mathrm{acc}(f)+2S_{n}^\mathrm{loc}+S_{n}^\mathrm{sn}+S_{n}^\mathrm{omn}}{L^2}
\left[1+\left(\frac{2Lf}{0.41 c}\right)^2\right],
\label{eq:sens}
\end{eqnarray}
where $L=2.5\times 10^{6}km$ is the arm length among satellites, and the noise
$S_{n}^\mathrm{acc}(f)$, $S_{n}^\mathrm{loc}$,
$S_{n}^\mathrm{sn}$ and $S_{n}^\mathrm{omn}$ denote the low-frequency acceleration, local interferometer noise,
shot noise and other measurement noise, respectively.
They can be written as the following according to the experimental results from  LISA Pathfinder~\cite{Armano:2016bkm}
\begin{eqnarray}
	S_{n}^\mathrm{acc}(f) & = & \left\{9 \times 10^{-30}+3.24 \times 10^{-28}\left[\left(\frac{3\times10^{-5}~\mathrm{Hz}}{f}\right)^{10} \right. \right. \nonumber \\
	&  & \left. \left. + \left(\frac{10^{-4}~\mathrm{Hz}}{f}\right)^{2}\right]\right\}\frac{1}{(2\pi f)^4}\,\mathrm{{m^2\,Hz}^{-1}},
\end{eqnarray}
and the other noise expressions are of the following:
\begin{equation}
	\begin{split}
		&S_{n}^\mathrm{sn}= 7.92\times10^{-23}~\mathrm{{m}^2\,{Hz}^{-1}},\\
		&S_{n}^\mathrm{omn}=4.00\times10^{-24}~\mathrm{{m}^2\,{Hz}^{-1}},\\
		&S_{n}^\mathrm{loc}= 2.89\times10^{-24}~\mathrm{{m}^2\,{Hz}^{-1}}.
	\end{split}
\end{equation}

%

\providecommand{\href}[2]{#2}\begingroup\raggedright

\end{document}